\documentclass{acm_proc_article-sp}

\usepackage[vlined,ruled,linesnumbered]{algorithm2e}
\usepackage{multirow}
\usepackage{floatflt}
\usepackage{amsfonts}
\usepackage{bbold}

\usepackage{graphicx}
\usepackage{amsmath}
\usepackage[tight,footnotesize]{subfigure}
\usepackage[caption=false,font=footnotesize]{subfig}
\usepackage{url}
\usepackage{lscape}

\newtheorem{definition}{Definition}
\newtheorem{example}{Example}

\begin{document}

\title{Cats\&Co: Categorical Time Series Coclustering\titlenote{Romain Guigour\`es was with Orange Labs when this work began.}}

%
%
%
%
%

\numberofauthors{4} 
%
\author{
%
%
\alignauthor
Dominique Gay\\
       \affaddr{Orange Labs}\\
\alignauthor
Romain Guigour\`es\\
       \affaddr{Zalando}\\
\and
\alignauthor Marc Boull\'e\\
       \affaddr{Orange Labs}\\
\alignauthor Fabrice Cl\'erot\\
       \affaddr{Orange Labs}\\
}

\date{24 september 2014}

\maketitle
\begin{abstract}
We suggest a novel method of clustering and exploratory analysis of temporal event sequences data (also known as categorical time series) based on three-dimensional data grid models. A data set of temporal event sequences can be represented as  a data set of three-dimensional points, each point is defined by three variables: a sequence identifier, a time value and an event value. Instantiating data grid models to the 3D-points turns the problem into 3D-coclustering.

The sequences are partitioned into clusters, the time variable is discretized into intervals and the events are partitioned into clusters. The cross-product of the univariate partitions forms a multivariate partition of the representation space, i.e., a grid of cells and it also represents a nonparametric estimator of the joint distribution of the sequences, time and events dimensions. Thus, the sequences are grouped together because they have similar joint distribution of time and events, i.e., similar distribution of events along the time dimension. The best data grid is computed using a parameter-free Bayesian model selection approach. We also suggest several criteria for exploiting the resulting grid through agglomerative hierarchies, for interpreting the clusters of sequences and characterizing their components through insightful visualizations. Extensive experiments on both synthetic and real-world data sets demonstrate that data grid models are efficient, effective and discover meaningful underlying patterns of categorical time series data. 
\end{abstract}

\category{H.2.8}{Database management}{Database applications}[Data mining]


\keywords{Clustering, Categorical Time Series, Parameter-free mining} 

\section{Introduction}
\label{sec:introduction}
Mining data with temporal information is a key challenge in the Knowledge Discovery process. Temporal data is complex given that an object is described by one or more sequences of time-ordered elements or events. Depending on the nature of the temporal events (categorical or numerical, time-points or time intervals), classical data mining techniques like pattern mining, clustering, classification have been instantiated for temporal data~\cite{Mor07}.

Here, we focus on categorical times series (cats) data (i.e., time-points event sequences data), where each event of a sequence is annotated by a time value $t$. Mining cats data is useful in many application domains, e.g., \cite{PBR+11} explore Electronic Medical Records data to find frequent temporal pattern of ICD codes across patients; \cite{MPT+08} look for frequent user behaviors in unexpected time periods from web logs; in social science domain, \cite{MGR+08} group individuals with similar life courses.
In the literature, a lot of the efforts have been dedicated to pattern mining in cats data, (e.g., frequent temporally-annotated sequence mining in~\cite{GNP06}) whereas summarizing through clustering such data has received less attention (see further related work discussed in section~\ref{sec:related}). Indeed, most of the clustering techniques for sequential data are dedicated to sequences without time annotations, i.e., only the placement or the sequentiality of events is relevant -- like in biological data, one of the most popular applications of sequence clustering.
%

In this paper, we suggest a methodology for clustering and exploratory analysis of cats data. From a domain expert point of view, a clustering of cats data should hold the following features:
\begin{enumerate}
\item \emph{Global picture:} the clustering technique should propose a global picture/summary of the underlying data structure and show the evolution of clusters of cats along the time dimension.
\item \emph{Local pattern detection:} the clustering technique should also highlight local patterns, e.g., combination of groups of events and time segments that characterize a particular cluster of cats w.r.t. the whole cats data.
\item \emph{Balancing generality and accuracy:} the resulting clustering should propose a valuable trade-off between generality (i.e., a concise summary) and accuracy the description of input data.
\item \emph{Parameter-free:} To facilitate the task of the analyst, computing the clustering should not involve parameter tuning.
\item \emph{Exploration abilities:} The whole methodology should take into account the expert needs and allow him to explore the resulting clustering w.r.t. each data dimension and at various granularities.
\end{enumerate}

To the best our knowledge, there exists no clustering technique for cats having all these properties. The methodology we suggest fulfills all the above requirements. The originality of our approach is that the cats data clustering problem is seen as a three-dimensional co-clustering problem. The three dimensions (or variables) are sequence identifiers, time and event.

\noindent \textbf{Roadmap:} In section~\ref{sec:method}, we suggest Khiops Co-clustering (\textsc{khc}), a 3D co-clustering method for categorical time series based on data grid models~\cite{Bou10}. \textsc{khc} aims at simultaneously partitioning the sequence identifiers into clusters, discretizing the time into intervals and partitioning the events into clusters by optimizing a Bayesian criterion that bets on a trade-off between the accuracy and robustness of the data grid model. The optimal grid is reached using an user parameter-free Bayesian selection method. In section~\ref{sec:exploitation}, we show how to exploit the resulting grid at various granularities by means of several criteria derived from the optimization criterion and information-theoretic measures. Section~\ref{sec:experiments} reports the experimental validation of our contributions on both synthetic and real-world data sets. We discussed further related work in section~\ref{sec:related} before concluding.

\begin{figure}[tbp!]%
\centering
\includegraphics[width=.9\columnwidth]{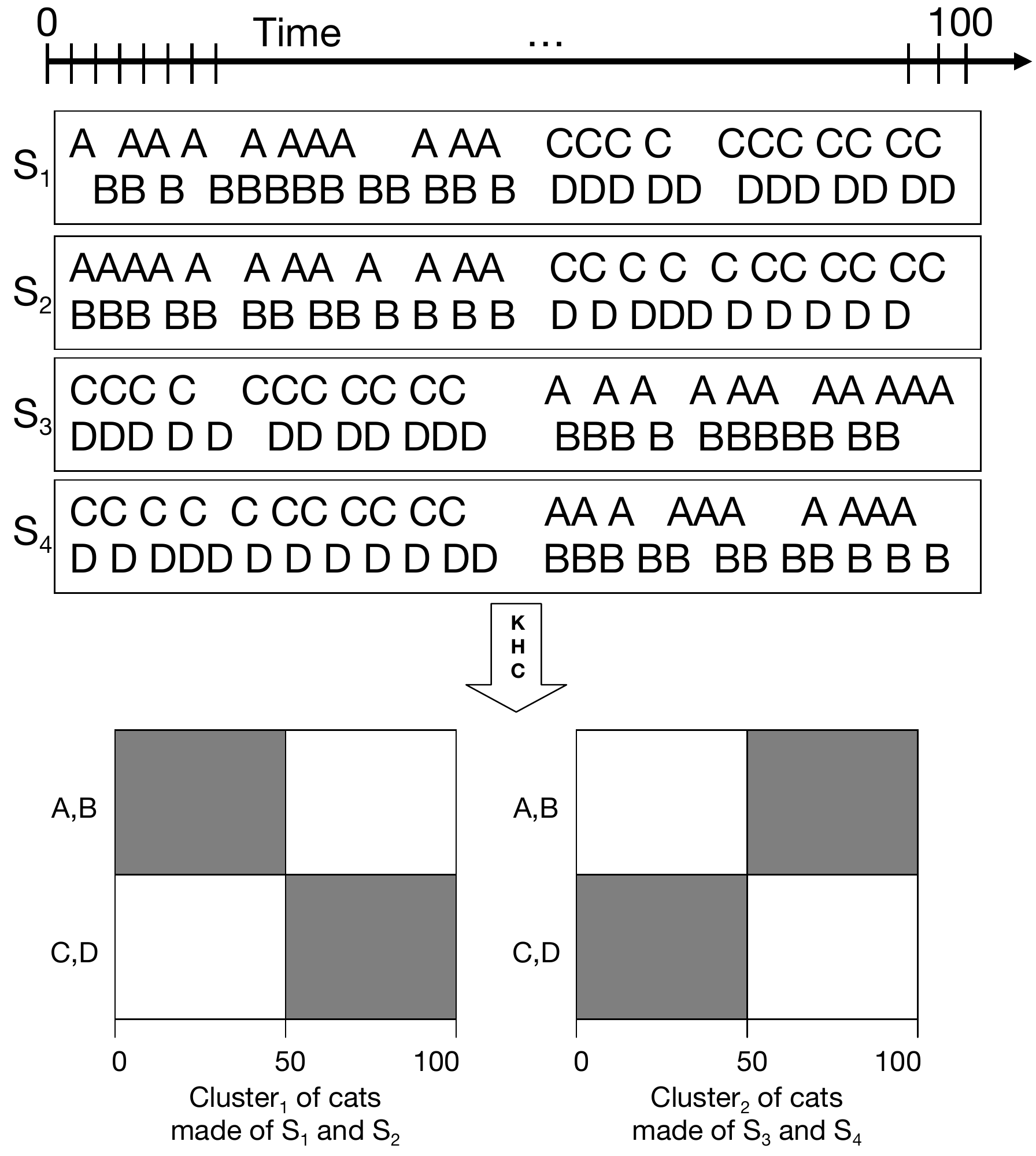}%
\caption{Visualization of results of \textsc{khc} on a toy sample data of 4 cats $\{S_1,S_2,S_3,S_4\}$, using 4 events $\{A,B,C,D\}$ over a $[0;100]$ timeline. The clustering highlight two clusters of cats: the first one composed of cats, with events $A$ and $B$ in time interval $[0;50]$ and with events $C$ and $D$ in time interval $]50;100]$; the second cluster shows an opposite behavior.}%
\label{fig:example}
\end{figure}

\begin{example}[From cats to data grid models]
Let us consider a toy example of cats data, made of 4 cats $S=\{S_1,S_2,S_3,S_4\}$, 4 events $E=\{A,B,C,D\}$ and a $T=[0;100]$ timeline. Figure~\ref{fig:example} shows the input cats data and the resulting 3D-coclustering which we split into two 2D ($T\times E$) coclustering, following the two clusters of cats we found.\\
Cats $S_1$ and $S_2$ are grouped together because they have similar joint distribution of time and events; in other words, they have similar distribution of events along the time dimension.\\
Cats $S_3$ and $S_4$ also have similar event distribution along the time, but their time-event distribution is clearly different from the distribution of $S_1$ and $S_2$, therefore they belong to a different cluster.\\
Two time segments are found, $T_1=[0;50]$ and $T_2=]50;100]$, which correspond to the two different regimes of events: $A,B$ on $T_1$ for $S_{1,2}$ then $C,D$ on $T_2$; and the opposite behavior for $S_{3,4}$.\\
\end{example}

%
%
\section{Categorical time series and data grid models}
\label{sec:method}

\begin{table*}[htbp!]
\centering
\caption{Notations and definitions}
\begin{tabular}{ll}
\hline
Notations & Defintions\\
\hline
$S,T,E,\mathcal{D}$ & cats identifiers variable, time variable, event variable of data $\mathcal{D}$\\
$n$ & number of sequences\\
$a$ & number of events in $E$\\
$N$ & number of points in $D$\\
$k_S$ (resp. $k_E$, $k_T$) & number of clusters of sequences (resp. clusters of events, time intervals)\\
$k=k_Sk_Ek_T$ & the number of cells of the grid\\
$N_{i_S}$ & cumulated number of points of cluster $i_S$ of sequences \\
$N_{j_T}$ & cumulated number of points in time interval $j_T$\\
$N_{i_E}$ & cumulated number of points of cluster $i_E$ of events\\
$N_{i_Sj_Ti_E}$ & cumulated number of points of the grid cell $(i_S,j_T,i_E)$)\\
$n_{i_S}$ (resp. $n_{i_E}$) & the number of sequences in cluster $i_S$ (resp. event values in cluster $i_E$)\\
$n^S_i$ (resp. $n^E_i$) & number of points in sequence $i$ (resp. with event value $i$)\\
\hline
\end{tabular}
\label{tab:notations}
\end{table*}

\noindent \textbf{Context and notations.} A temporal event sequence $s_i$ (or a categorical time series, say cats) of length $k_i>0$ is a time-ordered finite set of observations:
$$s_i = \langle (t_{i_1}, e_{i_1}), (t_{i_2}, e_{i_2}), \ldots , (t_{i_{k_i}} , e_{i_{k_i}})\rangle$$
such that $\forall j, 1\leq j\leq i_{k_i}, t_j\in\mathbb{R}+$, $e_j\in E$ and $E$ is a non-ordered set of categorical events. A cats data set is simply a set of such defined cats $\mathcal{D}=\{s_1,\ldots ,s_n\}$. We represent $D$ as a three-dimensional data set, i.e., with three variables (two categorical and one numerical variable): $S$ for the sequence (cats) id variable, $T$ for the time variable and $E$ for the event variable. In the following, an object $(s,t,e) \in \mathcal{D}$ is called a point of the data set and $N$ is the total number of points in $D$.\\
This general definition of cats data allows different size of cats and does not force the time stamps to be ``aligned'' for all the cats.


%
\subsection{Data grid models}
This 3D representation is suitable for co-clustering through data grid models~\cite{Bou10}. To make the paper self-contained, we recall the main features of the generic data grid model approach and describe its instantiation to cats data.

A data grid provides a piecewise constant joint density estimation of the input variables. Instantiating data grid models to the cats data, the goal is to simultaneously partition the categorical variables (sequence ids and events) into clusters and to discretize the numerical variable (time). The result is a 3D grid whose cells (say coclusters) are defined by a group of sequence ids, a group of events and a time interval. Notice that in all rigor, we are working only with partitions of variable value sets. However, to simplify the discussion we will sometime use a slightly incorrect formulation by mentioning a ``partition of a variable'' and a ``partitioned variable''.

In order to choose the ``best'' data grid model $M^{\ast}$ (given the data) from the model space $\mathcal{M}$, we use a Bayesian Maximum A Posteriori (MAP) approach. We explore the model space while minimizing a Bayesian criterion, called $cost$. The cost criterion bets on a trade-off between the accuracy and the robustness of the model and is defined as follows:
\begin{align}
cost(M) = -\log(\underbrace{p(M \mid D)}_{\textrm{posterior}}) \propto  -\log(\underbrace{p(M)}_{\textrm{prior}} \times \underbrace{p(D \mid M)}_{\textrm{likelihood}})\nonumber
\end{align}

We now define the model space $\mathcal{M}$ which consists of a family of cats data co-clustering models, based on clusters of cats ids, time intervals, clusters of events and a multinomial distribution of all the points on the cells of the resulting data grid.
\begin{definition}[Cats data grid models]
\label{def:model}
A cats data grid coclustering model is defined by:
\begin{itemize}
\item a number of clusters of cats ids,
\item a number of intervals for the time variable,
\item a number of clusters of events,
\item the repartition of the cats ids into the clusters of cats,
\item the repartition of the events into the clusters of events,
\item the distribution of the points of the cats data on the cells of the data grid,
\item for each cluster of cats (resp. of events), the distribution of the points that belongs to the cluster on the cats (resp. events) of the cluster.
\end{itemize}
\end{definition}
Boull\'e~\cite{Bou10} has shown that one can obtain an exact analytic expression of the $cost$ criterion if one consider a data-dependent hierarchical prior (on the parameters of a data grid model, see definition~\ref{def:model}) that is uniform at each stage of the hierarchy. Notice that it does not mean that the prior is uniform, thus in our case, the MAP approach is different from a simple likelihood maximization. The $cost$ criterion is then defined as follows:

\begin{definition}[cost: data grid evaluation]
\label{def:cost}
A data grid model for cats co-clustering is optimal if the value of the following $cost$ criterion is minimal:
\begin{flalign}
cost(M) & =\nonumber \\
& \log n + \log a + \log N + \log B(n,k_S) + \log B(a,k_E)\\
& + \log \binom{N+k-1}{k-1} \\
& + \sum_{i_S=1}^{k_S} \log \binom{N_{i_S}+n_{i_S}-1}{n_{i_S}-1}\\
& + \sum_{i_E=1}^{k_E} \log \binom{N_{i_E}+n_{i_E}-1}{n_{i_E}-1}\\
& + \log N! - \sum_{i_S=1}^{k_S} \sum_{j_T=1}^{k_T} \sum_{i_E=1}^{k_E} \log N_{i_Sj_Ti_E}!\\
& + \sum_{i_S=1}^{k_S} \log N_{i_S}! - \sum_{i=1}^{n} \log n^S_i!\\
& + \sum_{i_E=1}^{k_E} \log N_{i_E}! - \sum_{i=1}^{a} \log n^E_i!\\
& + \sum_{j_T=1}^{k_T} \log N_{j_T}!
\end{flalign}
%
where $B(n,k_S)$ is the number of partitions of $n$ elements into $k_S$ subsets and $B(a,k_E)$ is defined in a similar way.
\end{definition}

The first four lines stand for the \emph{a priori} probability of the grid model and constitute the regularization term of the model: the first line corresponds to the a priori term for the choice of the number of clusters for $S$ and $E$, the number of intervals for $T$ and the choice of partition of $S$ and $E$ into value groups. The second line represents the specification of the distribution of the $N$ points on the $k$ cells of the grid. The third line corresponds to the specification of the distribution of the points of each cluster of cats on the cats ids. The fourth line specifies the similar distribution for the events.\\
The last four lines stand for the likelihood of data given the model: the fifth line corresponds the likelihood  of the distribution of the points in the cells using a multinomial term. The sixth (resp. seventh) line is the likelihood of cats ids (resp. event values) locally to each cluster of cats (resp. events). The last line stands for the likelihood of ranks locally to each time interval.

The intuition behind the trade-off between the a priori (regularization) terms and the likelihood terms is as follows: complex models (with many clusters of cats and/or events and/or many time intervals) are penalized whereas models that are closest to the data are preferred. The extreme case where we have at most one point per cell will maximize the likelihood but we will get a very low a priori probability of the grid model, thus a high $cost$ value. The other side case, i.e., the null model, is when we have only one cell: we have high prior probability but very low likelihood, thus high $cost$ value. Grids with low $cost$ value indicate a high a posteriori probability $p(M \mid D)$ and are those of interest because they achieve a balanced trade-off between accuracy and generality. In terms of information theory, negative logarithm of probabilities can also be interpreted as code length: here, according to the Minimum Description Length principle (MDL), the $cost$ criterion can be interpreted as the code length of the grid model plus the code length of the data given the grid model; and a low $cost$ value also means a high compression of the data using grid model $M$.


\subsection{Data grid optimization}
\noindent \textbf{Optimization algorithm.} The optimization of data grid is a combinatorial problem: the number of possible partitions of $n$ cats is equal to the Bell number $B(n) = \frac{1}{e}\sum_{k=1}^{\infty}\frac{k^n}{k!}$ (we have a similar number for the event dimension $E$) and the number of discretizations of $N$ values is $2^N$. Obviously, an exhaustive search is unfeasible and as far as we know, there is no tractable optimal algorithm. Therefore the $cost$ criterion is optimized using a greedy bottom-up strategy whose main principle is described in pseudo-code Algorithm~\ref{algo:gbum}. We start with the finest grained data grid, that is made of the finest possible univariate partitions (of $S$, $T$ and $E$), i.e., based on single value intervals or clusters. Then, we evaluate all merges between clusters of sequence ids, clusters of events and adjacent time intervals and perform the best merge if the $cost$ criterion decreases after the merge. We iterate until there is no more improvement of the $cost$ criterion.
%

\begin{algorithm}

  \caption{\textsc{khc}: Cats data grid}
   \label{algo:gbum}
\SetKwInOut{Input}{Input}
\SetKwInOut{Output}{Output}
\SetFuncSty{textsc}
\SetArgSty{texttt}
\SetDataSty{texttt}
\SetKwComment{Comment}{}{} 

\Input{$M$ Initial data grid solution }
\Output{$M^{\ast}, cost(M^{\ast})\leq cost(M)$ final data grid solution with improved $cost$}
%

$M^{\ast} \leftarrow M$\;\nllabel{algo:gbum:initgrid}

\While{improved data grid solution}{ \nllabel{algo:gbum:stopping}
	$M' \leftarrow M^{\ast}$\; \nllabel{algo:gbum:inittgrid}
	\ForAll{Merge $m$ between two clusters of $S$ or $E$ or two intervals of $T$}{
		$M^+ \leftarrow M^{\ast} + m$ \Comment*[r]{//consider merge $m$ for grid $M^{\ast}$}
		\If{$cost(M^+) < cost(M')$}{
			$M' \leftarrow M^+$\;
		}
	}
	\If{$cost(M') < cost(M^{\ast})$}{
		$M^{\ast} \leftarrow M'$ \Comment*[r]{// Improved grid solution}
	}
}
\Return{$M^{\ast}$}
\end{algorithm}

A straightforward implementation of the greedy heuristic remains a hard problem since each evaluation of the $cost$ criterion for a grid $M$ requires $O(naN)$ time, given that the initial finest grid is made of up to $n\times a\times N$ cells (where $n$ is the number of cats ids, $a$ the number of events ($|E|$) and $N$ the number of points in $D$). Furthermore, each step of algorithm~\ref{algo:gbum} requires $O(n^2)$ (resp. $O(a^2)$, $O(N)$) evaluations of merges of clusters of cats ids (resp. clusters of events, time intervals); and there are at most $O(n+a+N)$ steps from the finest grained model to the null model. The overall time complexity is bounded by $O(naN(n^2+a^2+N)(n+a+N))$. In~\cite{Bou10}, it has been shown that further optimizations allow to reduce the time complexity to $O(N\sqrt{N}\log N)$. Advanced optimizations combined with sophisticated algorithmic data structures mainly exploits \textit{(i)} the sparseness of the grid, \textit{(ii)} the additivity property of the $cost$ criterion and \textit{(iii)} starts from non-maximal grained grid models using pre and post-optimization heuristics:
\begin{itemize}
\item[\textit{(i)}] Cats data sets represented by 3D points are sparse. Among the $O(naN)$ cells of the grid, at most $N$ cells are non-empty. The contribution of empty cells to the $cost$ criterion in definition~\ref{def:cost} is null, thus each evaluation of a data grid may be performed in $O(N)$ time through advanced algorithmic data structures.
\item[\textit{(ii)}] The additivity of the $cost$ criterion stems from the data-dependent hierarchical prior of criterion. It means that it can be split in a hierarchy of components of the grid model: the variables ($S$, $T$ and $E$), then the parts (clusters or intervals) and finally cells. The additivity property allows to evaluate all merges between intervals or clusters in $O(N)$ time. Moreover, the sparseness of the data set ensures that the number of revaluations (after the best merge is performed) is small on average.
\item[\textit{(iii)}] Instead of starting from the finest grained grid, for tractability concern, the algorithm starts from grids with at most $O(\sqrt{N})$ clusters or intervals. Dedicated preprocessing and postprocessing heuristics are employed to locally improve the initial and final solutions produced by algorithm~\ref{algo:gbum}. In these heuristics, the $cost$ criterion is post-optimized alternatively for each variable while the partitions of the others are fixed, by moving values across clusters and moving interval boundaries for the time variable. 
\end{itemize}
The optimized version of algorithm~\ref{algo:gbum} is now time-efficient but may lead to a local optimum. To alleviate this concern, we use the Variable Neighborhood Search (VNS) meta-heuristic~\cite{HM01}. The main principle consists of multiple runs of the algorithms using various random initial solutions (we consider 10 rounds of initialization): it allows anytime optimization -- the more you optimize, the better the solution -- while not growing the overall time complexity of algorithm~\ref{algo:gbum}. Full details of the optimization techniques are available in~\cite{Bou10}.



%
\section{Exploiting the grid}
\label{sec:exploitation}

In some real-world large-scale case studies, the optimal grid $M^{\ast}$ resulting from the optimization algorithm \textsc{khc} is made of several hundreds of clusters of cats ids (or intervals and/or clusters of events), i.e. millions of cells, which is difficult to exploit and interpret. To alleviate this issue, we suggest a grid simplification method together with several criteria that allow us to choose the granularity of the grid for further analysis, to rank values in clusters and to gain insights in the data through meaningful visualizations.

\subsection{Data grid simplification}
\noindent \textbf{Dissimilarity index and grid structure simplification.}
We suggest a simplification method of the grid structure that iteratively merge clusters or adjacent intervals -- choosing the merge generating the least degradation of the grid quality. To this end, we introduce a dissimilarity index between clusters or intervals which characterize the impact of the merge on the $cost$ criterion.

\begin{definition}[Dissimilarity index]
Let $c_{.1}$ and $c_{.2}$ be two parts of a dimension partition of a grid model $M$ (i.e. two clusters of sequence ids or events or two adjacent intervals). Let $M_{c_{.1}\cup c_{.2}}$ be the grid after merging $c_{.1}$ and $c_{.2}$. The dissimilarity $\Delta(c_{.1},c_{.2})$ between the two parts $c_{.1}$ and $c_{.2}$ is defined as the difference of $cost$ before and after the merge:
\begin{align}
\Delta(c_{.1},c_{.2}) = cost(M_{c_{.1}\cup c_{.2}}) - cost(M) \nonumber
\end{align}
\end{definition}

When merging clusters that minimize $\Delta$, we obtain the sub-optimal grid $M'$ (with a coarser grain, i.e. simplified) with minimal $cost$ degradation, thus with minimal information loss w.r.t. the grid $M$ before merging.\\
Performing the best merges w.r.t. $\Delta$ iteratively over the three partitioned variables without distinction, starting from $M^{\ast}$ until $M_{\emptyset}$, three agglomerative hierarchies are built and the end-user can stop at the chosen granularity that is necessary for the analysis while controlling either the number of clusters/cells or the information ratio kept in the model. The information ratio of the grid $M'$ is defined as follows:
\begin{align}
IR(M') = (cost(M')-cost(\mathcal{M_{\emptyset}})) / (cost(M^{\ast}) - cost(M_{\emptyset})) \nonumber
\end{align}

where $M_{\emptyset}$ is the null model (the grid where no dimension is partitioned).\\
Building the hierarchies from $M^{\ast}$ to $M_{\emptyset}$ for the partitioned variables $S$, $T$ and $E$ shows a quadratic time complexity w.r.t. the total number of parts of the partitioned variables of $M^{\ast}$. However, generally, \textsc{khc} has already done the hard work: the number of parts is small. In practice, the computational time for building the hierarchies is negligible compared with the optimization phase.\\

\subsection{Ranking cats and events}
\noindent \textbf{Typicality for ranking categorical values in a cluster.} When the chosen granularity is reached through agglomerative hierarchy, the number of clusters per categorical dimension (cats ids or events) decreases and mechanically the number of values per cluster increases. It could be useful to focus on the most representative values (cats ids or events) among thousands of values of a cluster. In order to rank values in a cluster, we define the typicality of a value as follows:
\begin{definition}[Typicality of a value in a cluster]
For a value $v$ in a cluster $c$ of the partition $X^M$ of dimension $X$ given the grid model $M$, the typicality of $v$ is defined as:
\begin{equation}
\begin{array}{l}
\tau (v,c) =\nonumber \\
\frac{1}{1-P_{X^M}(c)} \times \nonumber \\
\sum_{c_j\in X^M\atop c_j\neq c} P_{X^M}(c_j)(cost(M|c\setminus v,c_j\cup v)-cost(M)) \nonumber
\end{array}
\end{equation}
where $P_{X^M}(c)$ is the probability of having a point with its value in cluster $c$, $c\setminus v$ is the cluster $c$ from which we have removed value $v$, $c_j\cup v$ is the cluster $c_j$ to which we add value $v$ and $M|c\setminus v,c_j\cup v$ the grid model $M$ after the aforementioned modifications.
\end{definition}
Intuitively, the typicality evaluates the average impact in terms of $cost$ on the grid model quality of removing a value $v$ from its cluster $c$ and reassigning it to another cluster $c_j\neq c$. Thus, a value $v$ is representative (say typical) of a cluster $c$ if $v$ is ``close'' to $c$ and ``different in average'' from other clusters $c_j\neq c$.\\
%
%

\subsection{Insightful visualizations}
\noindent \textbf{Insightful visualizations with Mutual Information and Contrast.}
It is common to visualize 2D coclustering results using 2D frequency matrix or heat map. For 3D coclustering, it is useful to select a dimension of interest (in our case, sequence ids $S$) and then we are able to visualize the frequency matrix of the two other dimensions ($T$ and $E$) given a cluster $c$ of $S$. We also suggest two other insightful measures for coclusters to be visualized, namely, the Contribution to Mutual Information (CMI) and the Contrast -- providing additional valuable visual information inaccessible with only frequency representation. Notice that the contributed visualizations are also valid whatever the dimension of interest.

\begin{definition}[Mutual Information and Contribution]
For a cluster of cats ids $c_{i_S}$, the mutual information between two partitioned variables $T^{\pi_M}$ and $E^{\pi_M}$ (from the partition $\pi_M$ of time and event variables induced by the grid model $M$) is defined as:
\begin{align}
MI(T^{\pi_M};E^{\pi_M})=\sum_{i_1=1}^{i_1=k_{T}}\sum_{i_2=1}^{i_2=k_{E}}MI_{i_1i_2} \nonumber \\
\text{where} \quad MI_{i_1i_2}=p(c_{i_1i_2})\log \frac{p(c_{i_1i_2})}{p(c_{i_1})p(c_{i_2})} \nonumber
\end{align}
where $MI_{i_1i_2}$ represent the contribution of cell $c_{i_1i_2}$ to the mutual information.
\end{definition}
Thus, if $MI_{i_1i_2} > 0$ then $p(c_{i_1i_2}) > p(c_{i_1})p(c_{i_2})$ and we observe an excess of interaction between $c_{i_1}$ and $c_{i_2}$ located in cell $c_{i_1i_2}$ defined by time interval $T_{i_1}$ and group of events $E_{i_2}$. Conversely, if $MI_{i_1i_2} < 0$, then $p(c_{i_1i_2}) < p(c_{i_1})p(c_{i_2})$, and we observe a deficit of interactions in cell $c_{i_1i_2}$. Finally, if $MI_{i_1i_2} = 0$, then either $p(c_{i_1i_2})=0$ in which case the contribution to MI is null and there is no interaction or $p(c_{i_1i_2}) = p(c_{i_1})p(c_{i_2})$ and the quantity of interactions in $c_{i_1i_2}$ is that expected in case of independence between the partitioned variables.\\

%

\begin{definition}[Contrast]
%
The contrast between the two partitioned variables to be visualized $(T^{\pi_M},E^{\pi_M})$ considered jointly and $S^{\pi_M}$ is defined as:
\begin{align}
Contrast((T^{\pi_M},E^{\pi_M}),S^{\pi_M}) = \sum_{i_S=1}^{i_S=k_{S}}\sum_{i_1=1}^{i_1=k_{T}}\sum_{i_2=1}^{i_2=k_{E}}MI_{i_Si_1i_2} \nonumber \\
\text{where} \quad MI_{i_Si_1i_2}=p(c_{i_Si_1i_2})\log \frac{p(c_{i_Si_1i_2})}{p(c_{i_1i_2})p(c_{i_S})} \nonumber
\end{align}
\end{definition}
Again, the sign of $MI_{i_Si_1i_2}$ values explains what is contrasting in $c_{i_S}$ w.r.t. the set of all sequence ids from the view $(T^{\pi_M},E^{\pi_M})$. Positive and negative values both highlight the cells that characterize $c_{i_S}$ w.r.t. the set of all sequence ids; the former says that an excess of interaction is located in cell $c_{i_1i_2i_S}$, the latter highlights a deficit of interaction (a negative contrast); and $MI_{i_1i_2i_S}=0$ (or nearly) indicates no significant contrast.

While the visualization of CMI of the cells highlight valuable information that is local to a cluster of cats, the contrast is a global scope visualization. Both CMI and contrast bring complementary insights to exploit the summary provided by the grid. In our experiments, we show the added-value of those visualizations on both synthetic and real data sets.

\section{Experimental validation}
\label{sec:experiments}
Our grid-based co-clustering method \textsc{khc} and visualization tools are both available under the name \textsc{khiops} at \url{http://www.khiops.com}. In this section, to validate our contributions, we report the experimentations on both synthetic and real-world large-scale \textsc{dblp} data sets. These experiments are designed to answer the following questions:
\begin{enumerate}
\item Effectiveness: How successful is \textsc{khc} in co-clustering cats, i.e., finding meaningful clusters of cats ids and events and intervals of time ?
\item Efficiency / Sacalability: Considering computational time, how does \textsc{khc} scale w.r.t. the data size and characteristics (i.e., the number of points, cats ids, events, underlying pattern to be discovered and noise) ?
\item Knowledge and insights: What kind of insights do the resulting grid and the exploitation tools bring in our knowledge of the data ?
\end{enumerate}

\subsection{Synthetic data sets}
Let us consider two patterns $M_1$ and $M_2$ defined on the time domain $T=[0;1000] \subseteq \mathbb{R}^+$ and the set of events $E=\{a,b,\ldots,k,l\}$ such that table~\ref{tab:pattern} defines:
\begin{table}[htbp!]
\centering
\caption{Two synthetic patterns: definition.}
\begin{tabular}{l}
$M_1$\\
\hline
$t\in T^{M_1}_1=[0;250] \Rightarrow e\in E^{M_1}_1=\{a,b,c\}$\\
$t\in T^{M_1}_2=]250;500] \Rightarrow e\in E^{M_1}_2=\{d,e,f\}$\\
$t\in T^{M_1}_3=]500;750] \Rightarrow e\in E^{M_1}_3=\{g,h,i\}$\\
$t\in T^{M_1}_4=]750;1000] \Rightarrow e\in E^{M_1}_4=\{j,k,l\}$\\
%
\\
$M_2$\\
\hline
$t\in T^{M_2}_1=[0;100] \Rightarrow e\in E^{M_2}_1=\{j,k,l\}$\\
$t\in T^{M_2}_2=]100;400] \Rightarrow e\in E^{M_2}_2=\{g,h,i\}$\\
$t\in T^{M_2}_3=]400;600] \Rightarrow e\in E^{M_2}_3=\{d,e,f\}$\\
$t\in T^{M_2}_4=]600;1000] \Rightarrow e\in E^{M_2}_4=\{a,b,c\}$\\
\end{tabular}
\label{tab:pattern}
\end{table}

Let us consider 10 cats following pattern $M_1$ and 10 cats for pattern $M_2$ (we also did the experiments for $CM=$50 and 100 cats per pattern). We generate a data set $D$ of $N = 2^{20}$ points (i.e., $5.10^4$ points in average per cats). Each point is a triplet with a randomly chosen cats id (among 20), a random value $t$ (on $T$) and an event value $e$ generated according to the pattern $M_i$ related to the cats id, i.e. an event value randomly chosen in the set $M_i(t)$ (see Table~\ref{tab:pattern}). Furthermore, we consider several noisy versions of this data set at various noise level $\eta = \{0.1,0.2,0.3,0.4,0.5\}$: when generating a point, the probability that the event value fulfills the pattern $M_i$ definition is $p(e\in M_i(t))=1-\eta$ and $p(e\in \{E \setminus M_i(t)\})=\eta$.

\begin{figure}[htbp!]%
\centering
\begin{tabular}{c}
\includegraphics[width=.9\columnwidth]{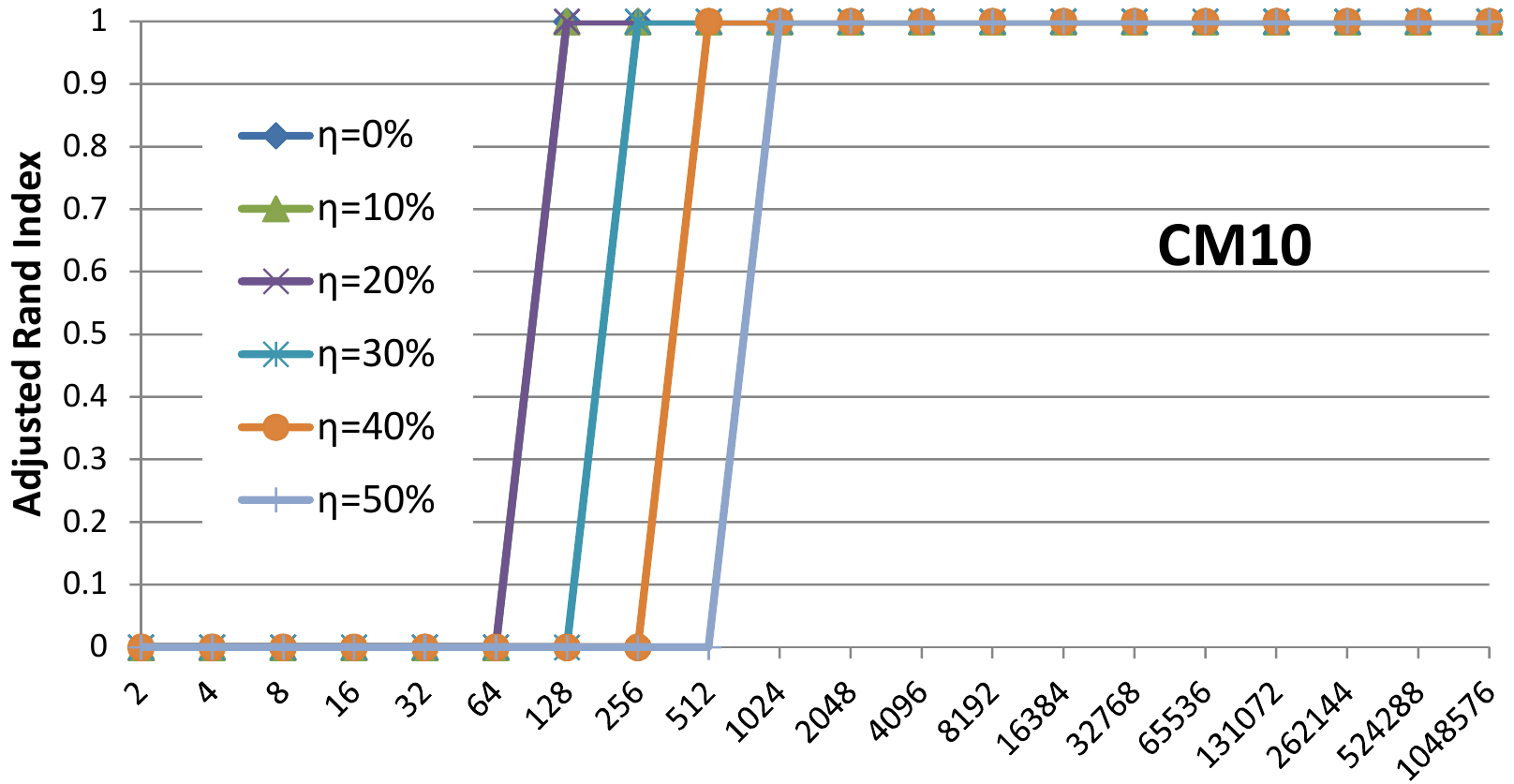}\\
\includegraphics[width=.9\columnwidth]{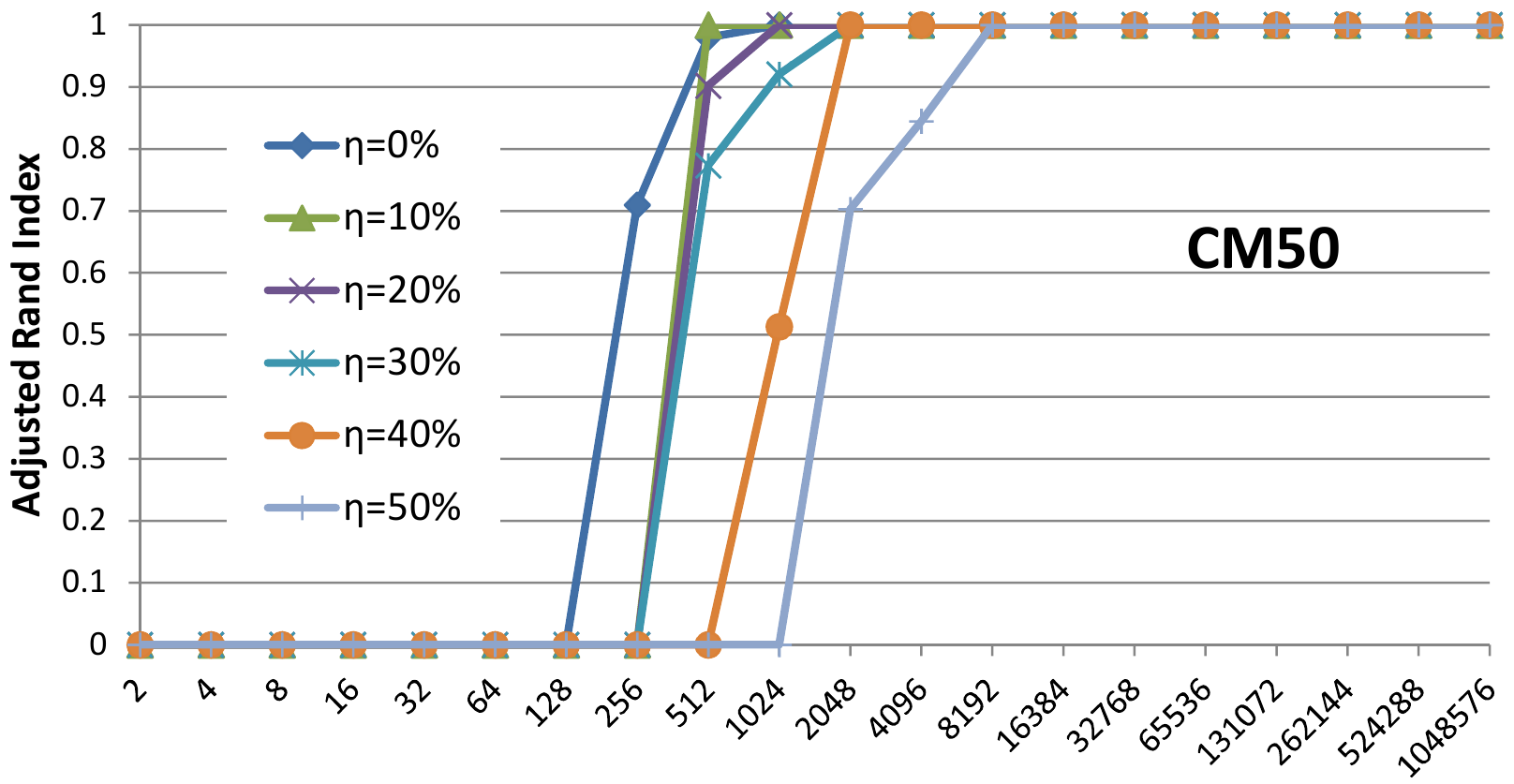}\\
\includegraphics[width=.9\columnwidth]{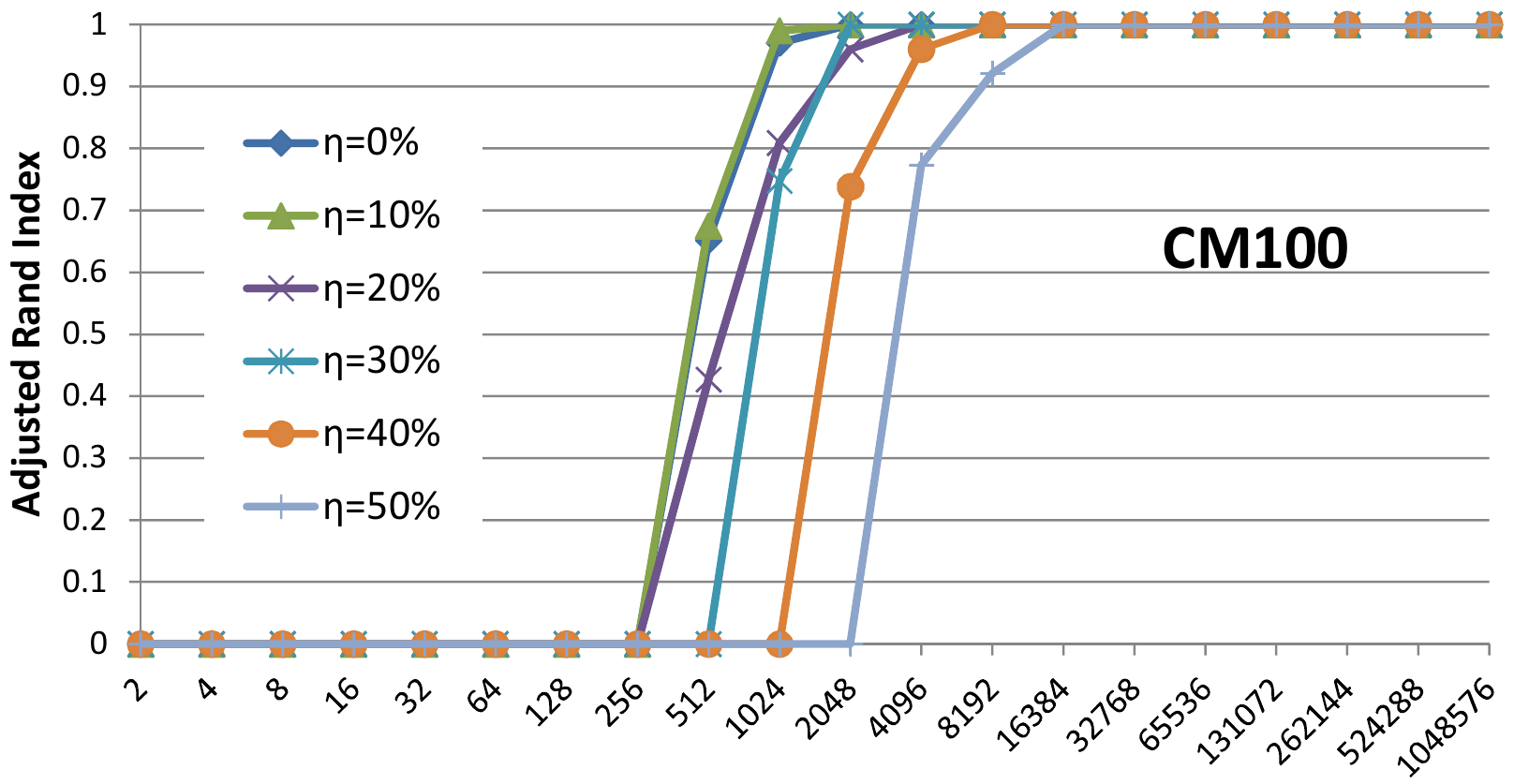}\\
\end{tabular}
\caption{Evolution of ARI for synthetic two-pattern cats data sets, for CM = 10, 50 and 100 cats per pattern and at various levels of noise w.r.t. number of points $N$.}%
\label{fig:adjrand}%
\end{figure}

We apply \textsc{khc} to subsets of $D$ of increasing sizes: $N$ varying from $2^1$ to $2^{20}$, i.e., overall \textsc{khc} is experimented through 360 various synthetic data sets. We compute the Adjusted Rand Index (ARI) for each grid generated by \textsc{khc} to evaluate the agreement between cats id clusters of the grid and the two underlying patterns.\\
The results are reported in Figure~\ref{fig:adjrand}. We observe that for small subsets of $D$, there is not enough points for \textsc{khc} to discover significant patterns: no cluster of cats id is found for $N \leq 64$ (i.e., in average 3 points per cats). For $CM=10$ (10 cats per pattern in figure~\ref{fig:adjrand}(a)), beyond $N=128$ points (6 points per cats in average), we have $ARI=1$ and the two underlying patterns are discovered. We see also that at noise level $\eta \leq 0.1$ , $N=128$ points are enough to find the patterns; then, more noise implies that more points are necessary to discover the patterns. Finally, increasing the number of points up to $2^{20}$ does not lead to over-fitting, $ARI=1$ and is stable. The same observations hold for $CM=50$ and $CM=100$: when the number of cats per pattern increases more points are needed.\\
Concerning the other variables (time $T$ and event $E$), generally speaking, when considering the increasing number of points, the true segmentation of time is discovered at the same step (or just before) and the true clustering of events is discovered first, i.e., just before the clustering of cats -- both remaining stable with increasing number of points in the data.
%

%
%

\begin{figure}[tbp!]%
\centering
\begin{tabular}{c}
\includegraphics[width=\columnwidth]{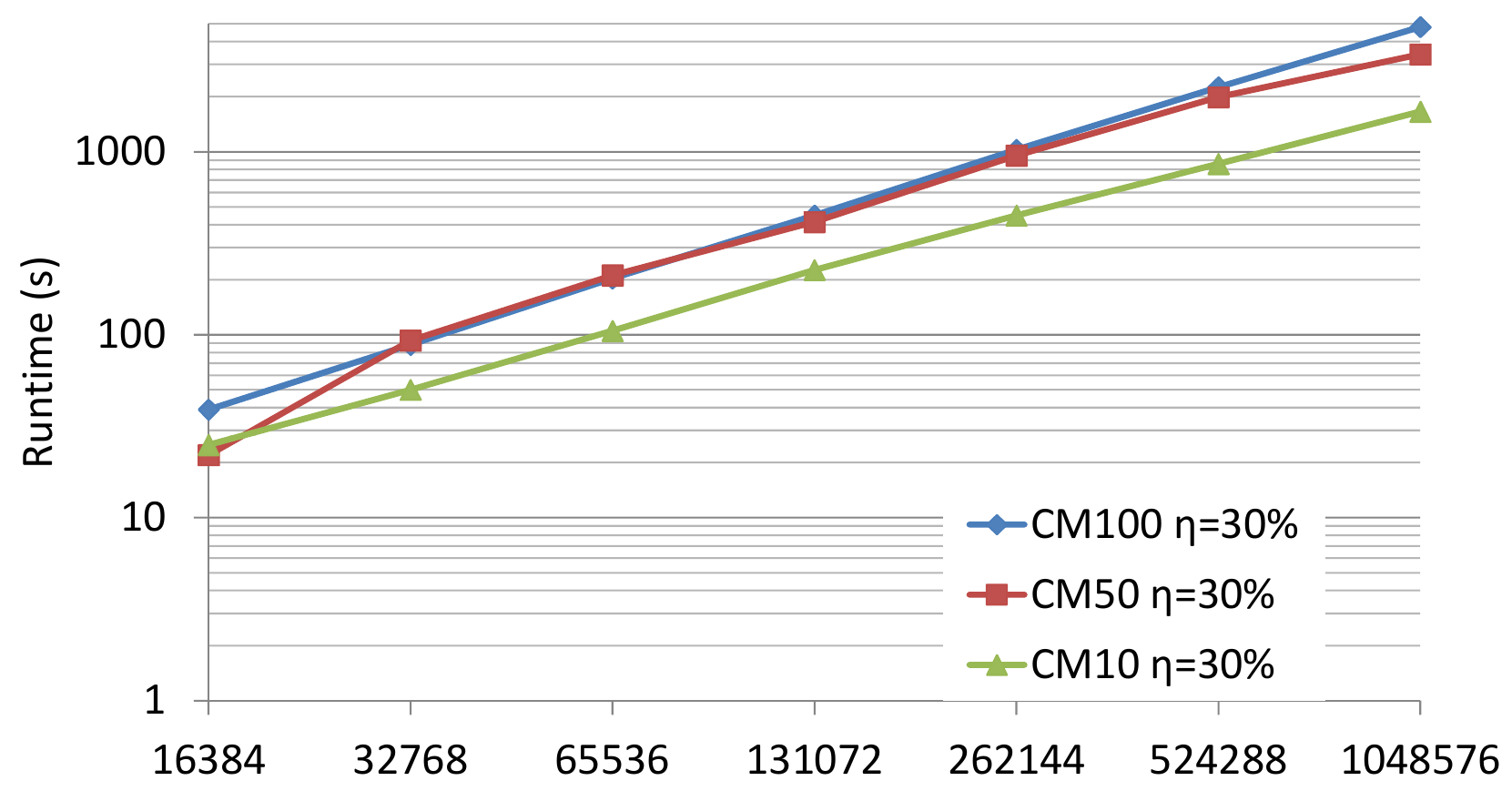}\\
\includegraphics[width=\columnwidth]{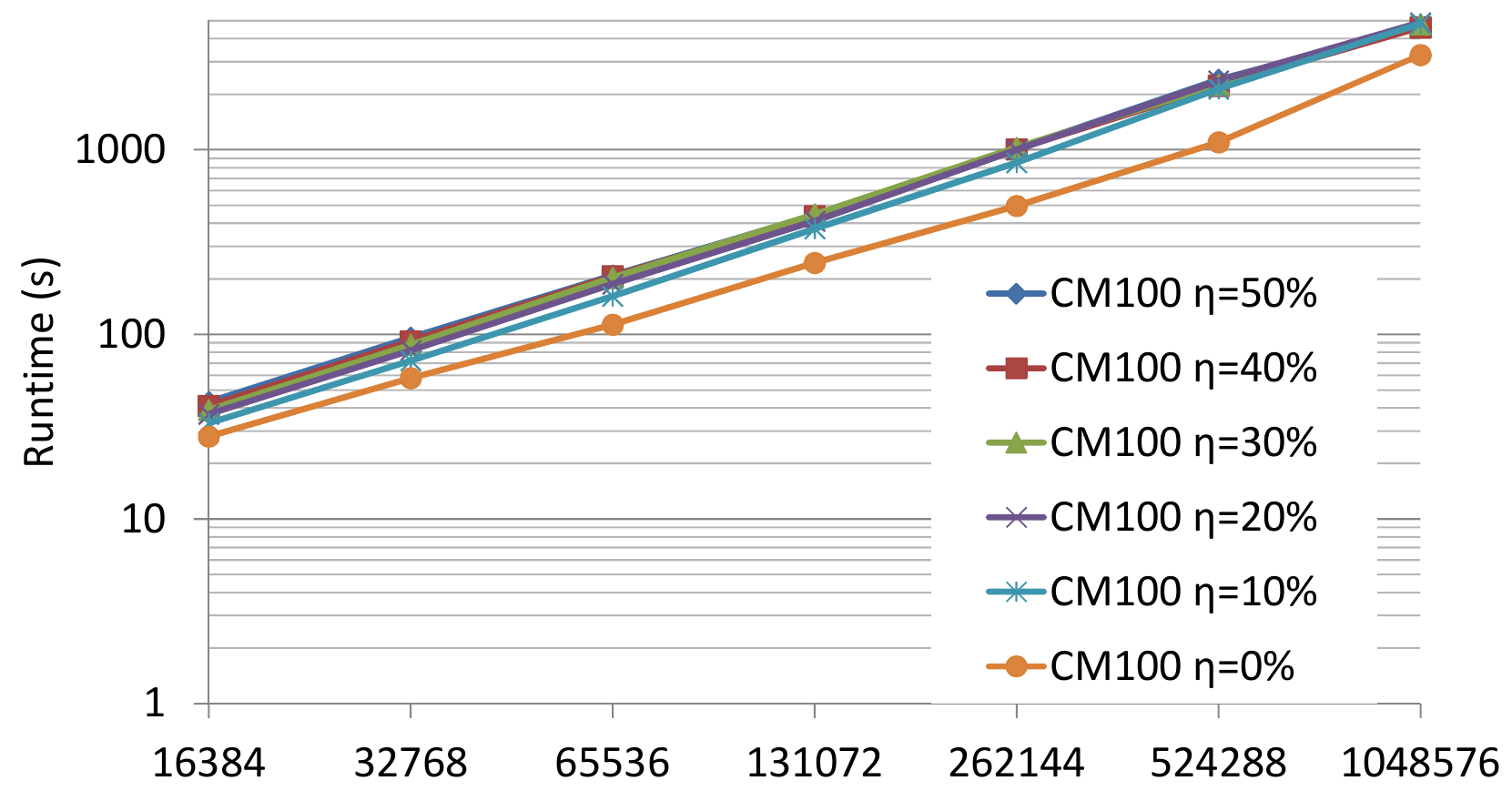}\\
\end{tabular}
\caption{Running time of \textsc{KHC} w.r.t. number of points ($N$), number of cats per pattern ($CM$) and noise level ($\eta$).}%
\label{fig:runtime}%
\end{figure}

\noindent \textbf{Running time.} Figure~\ref{fig:runtime} reports running time of \textsc{khc} on various versions of two-pattern data sets for $CM=10, 50, 100$ w.r.t. the number of points $N$. As expected, running time increases with the number of points in $D$ but also with $CM$ and $\eta$. For the most ``difficult'' data set, i.e., $N=2^{20}$, $CM=100$ (5200 points per cats in average) and $\eta=0.5$, \textsc{khc} finds the two underlying patterns in about 90 minutes: the difficulty comes from the time dimension (potentially $2^{20}$ different values).\\
A similar experiment has been led while considering integer values of time ($T = [0;1000] \subseteq \mathbb{N}^+$); in that case, \textsc{khc} finds the patterns faster, in 13 minutes. We have led other similar experiments with 5 and 10 underlying patterns to be discovered.
The main result is that more cluster patterns require more computational time and more points to be detected.

\begin{figure*}[tbp!]%
\centering
\begin{tabular}{ccc}
Frequency & CMI & Contrast\\
\includegraphics[width=.5\columnwidth]{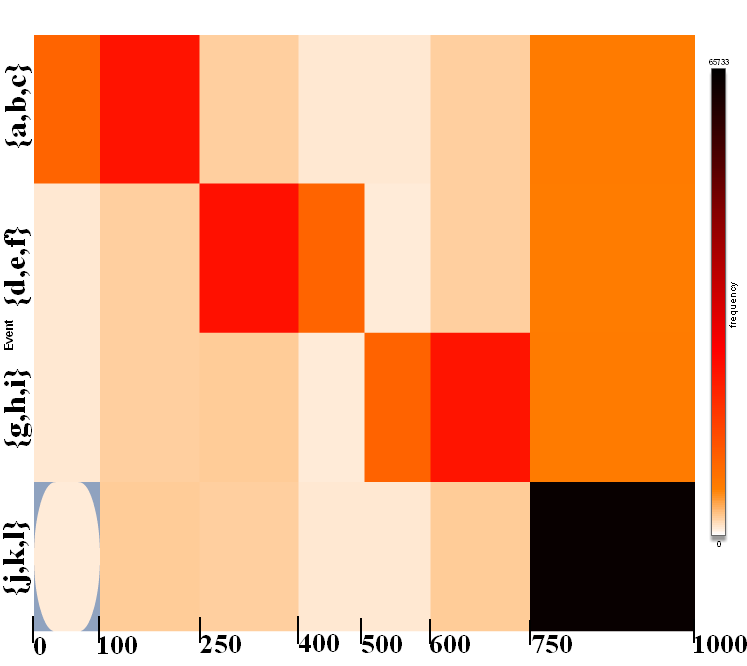}%
&
\includegraphics[width=.5\columnwidth]{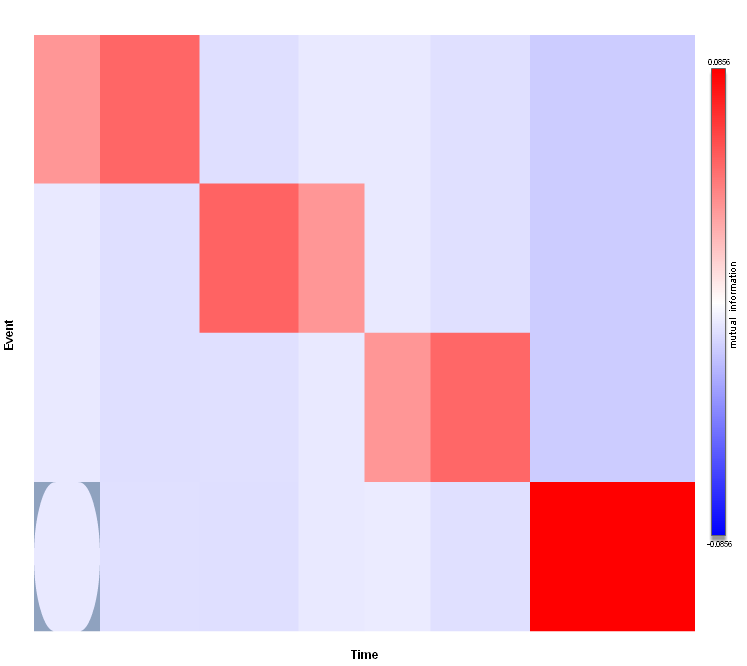}%
&
\includegraphics[width=.5\columnwidth]{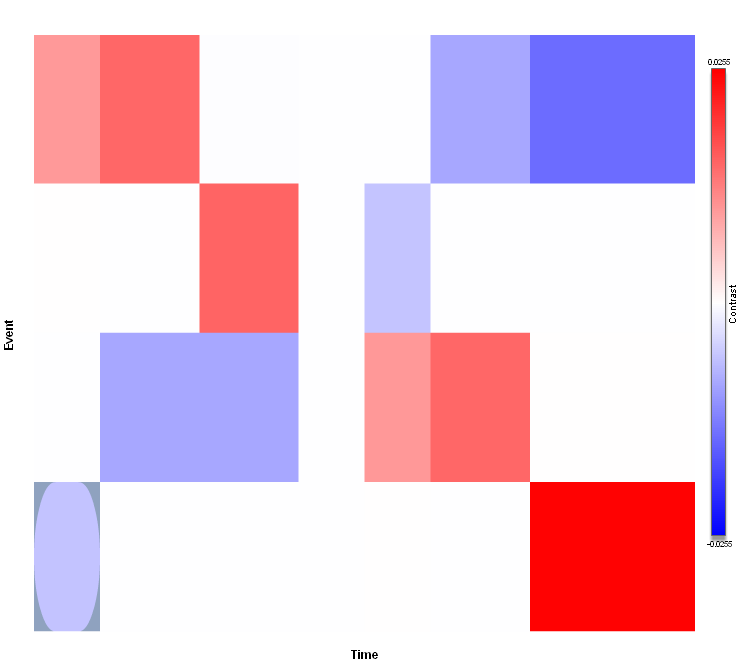}\\
$M_1$(a) & $M_1$(b) & $M_1$(c)\\%
\includegraphics[width=.5\columnwidth]{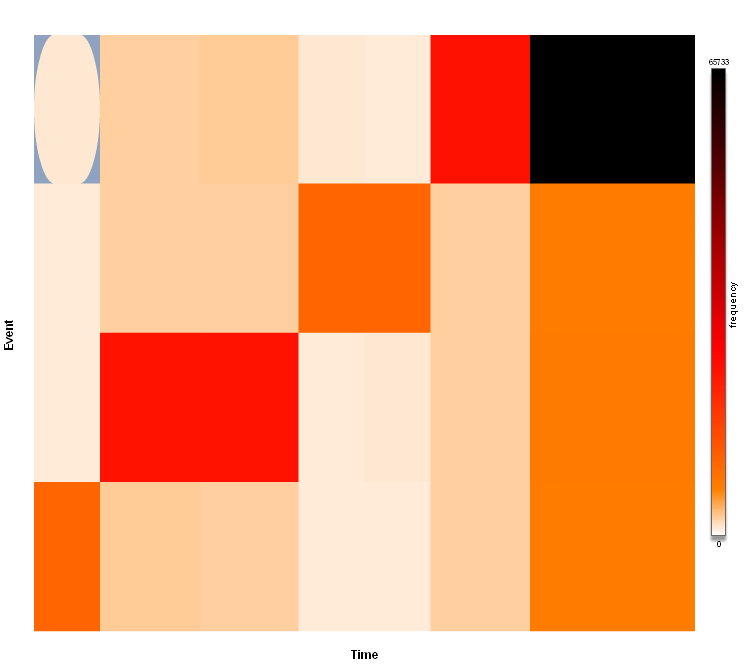}%
&
\includegraphics[width=.5\columnwidth]{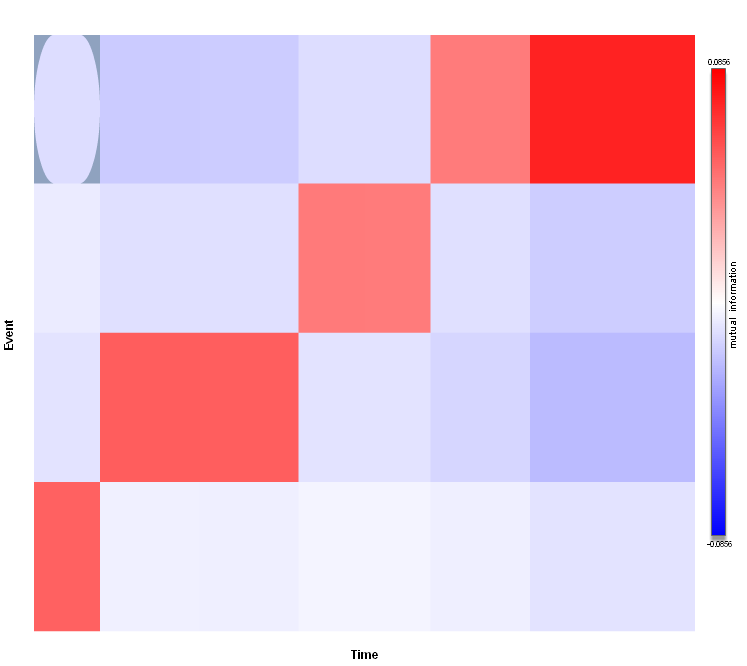}%
&
\includegraphics[width=.5\columnwidth]{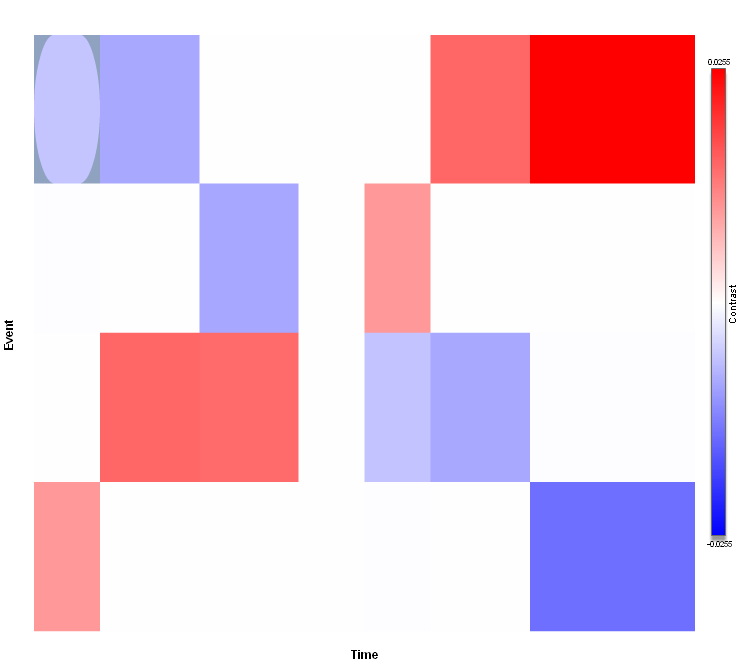}\\
$M_2$(a) & $M_2$(b) & $M_2$(c)\\%
\end{tabular}
\caption{$T\times E$ 2D-view: visualization of Frequency, Contribution to Mutual Information (CMI) and contrast for the two clusters of cats (corresponding to the two underlying patterns $M_1$ and $M_2$) found by \textsc{KHC}.
}%
\label{fig:2cats_vis}%
\end{figure*}

\noindent \textbf{Visualization and characterization of clusters.} Let us consider the 3D-grid obtained by \textsc{khc} on the two-pattern data set with $N=2^{20}$ points, $CM=10$ and $\eta=0.5$. Figure~\ref{fig:2cats_vis} shows 2D-views ($T\times E$) of each cluster of cats found by \textsc{khc}. Frequency, CMI and contrast visualizations bring different insights in the data and valuable information on the found clusters.\\
In figures~\ref{fig:2cats_vis}$M_1$(a) and $M_2$(a), for the frequency visualization (i.e., the number of points $N_{i_Sj_Ti_E}$ for a cell $i_Sj_Ti_E$), we already perceive the underlying patterns $M_1$ and $M_2$; however, the noise level $\eta=0.5$ degrades the visibility of patterns.\\
Figures~\ref{fig:2cats_vis}$M_1$(b) and $M_2$(b) bring to light and characterize cluster patterns $M_1$ and $M_2$. Red cells relate positive CMI, i.e., excess of interactions between $T$ and $E$ in a cell conditionally to current cats ids cluster -- that characterizes pattern cluster. Light blue cells stand for negative $MI_{i_1i_2}$ values, i.e. a slight deficit of interactions corresponding to noisy cells that are not significant to definition of the pattern cluster.\\ Finally, figures~\ref{fig:2cats_vis}$M_1$(c) and $M_2$(c) show the contrasting cells of each cats ids cluster: red for positive contrast, blue for negative contrast and white for no contrast. For example, let consider cluster pattern $M_1$: despite the noise level, white cell $([0;100],\{g,h,i\})$ is not characteristic of $M_1$ since probability of event group $\{g,h,i\}$ in time interval $[0;100]$ is not different from $M_1$ to $M_2$. Also, white cell $([401;500],\{d,e,f\})$ shows null contrast since it is common to the definition of the two underlying patterns (there is a similar distribution of points in the cell due to our data generator). Finally, for $M_1$, red cells an excess of interactions whereas blue cells indicates negative contrast (a deficit of interactions) -- that gives the mirror effect between figures~\ref{fig:2cats_vis}$M_1$(c) and $M_2$(c).

%


\subsection{Big pictures from DBLP bibliography}
In this section, we report the results of an exploratory analysis of the DBLP data set using our contributions. The DBLP Computer Science Bibliography~\cite{Ley09} records millions of publications (mainly from journals and conference proceedings) of Computer Science authors since 1936.\\
Let us consider a cats-like view of DBLP as a 3D point data set: $Author\times Year\times Event$, i.e. we consider author's sequence of publications over the years as cats data: \emph{Author} is the cats id, \emph{Event} is the name of the journal/proceedings/other where an author has published and \emph{Year} is the year he has published in the current \emph{event}. Duplicated points indicate that an author has published more than once in an event the same year (like e.g., (G. Alonso, 2009, EDBT) appearing twice in the data).
In this form, DBLP data (downloaded in august 2013; 2013 was still incomplete and 2014 referencing just began) contains more than 6.352 million points -- described by more than 1.297 million authors who have published in 6767 events from 1936 to 2014.

\begin{figure}[htbp!]%
\centering
\begin{tabular}{ccc}
\includegraphics[width=.3\columnwidth]{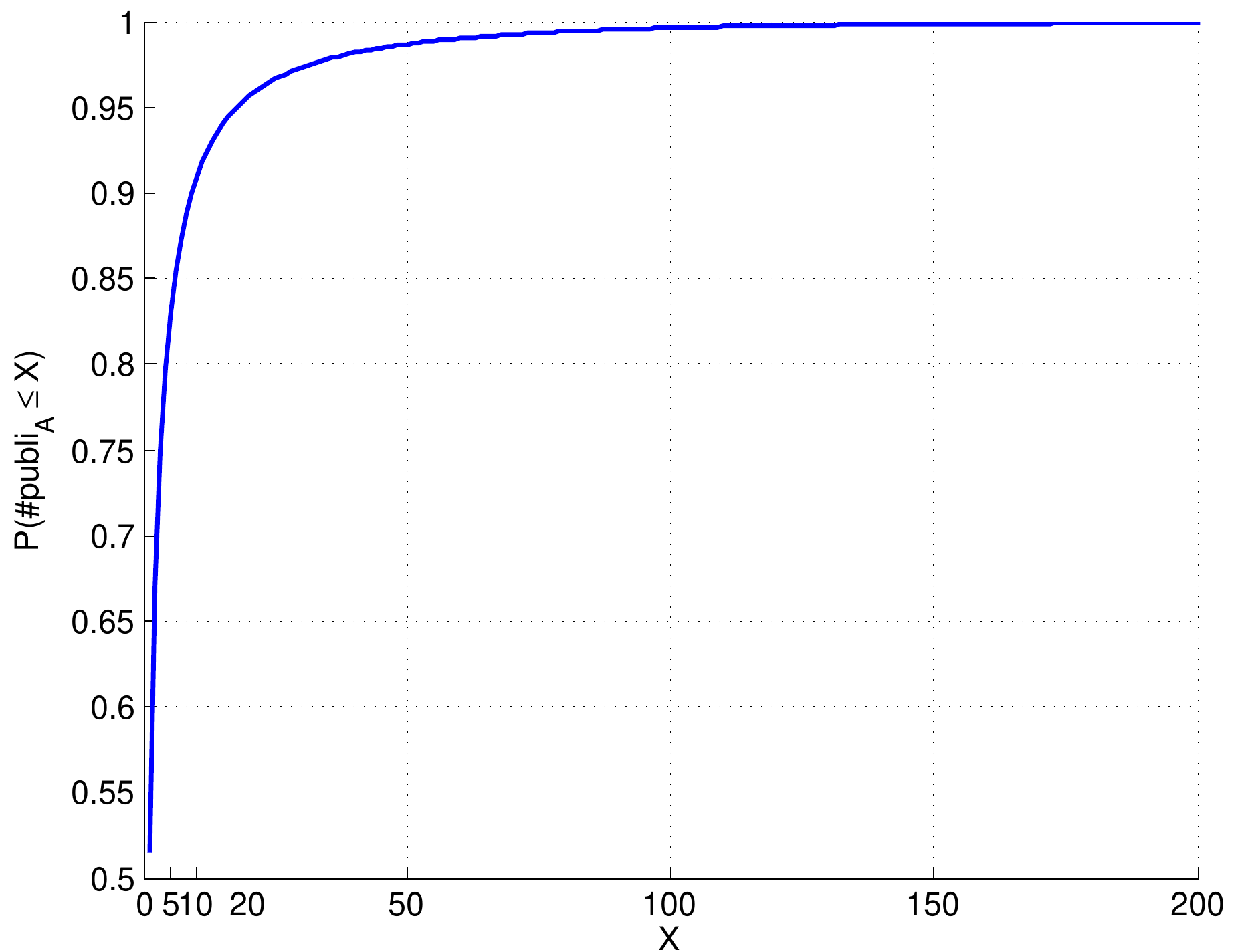}%
&
\includegraphics[width=.28\columnwidth]{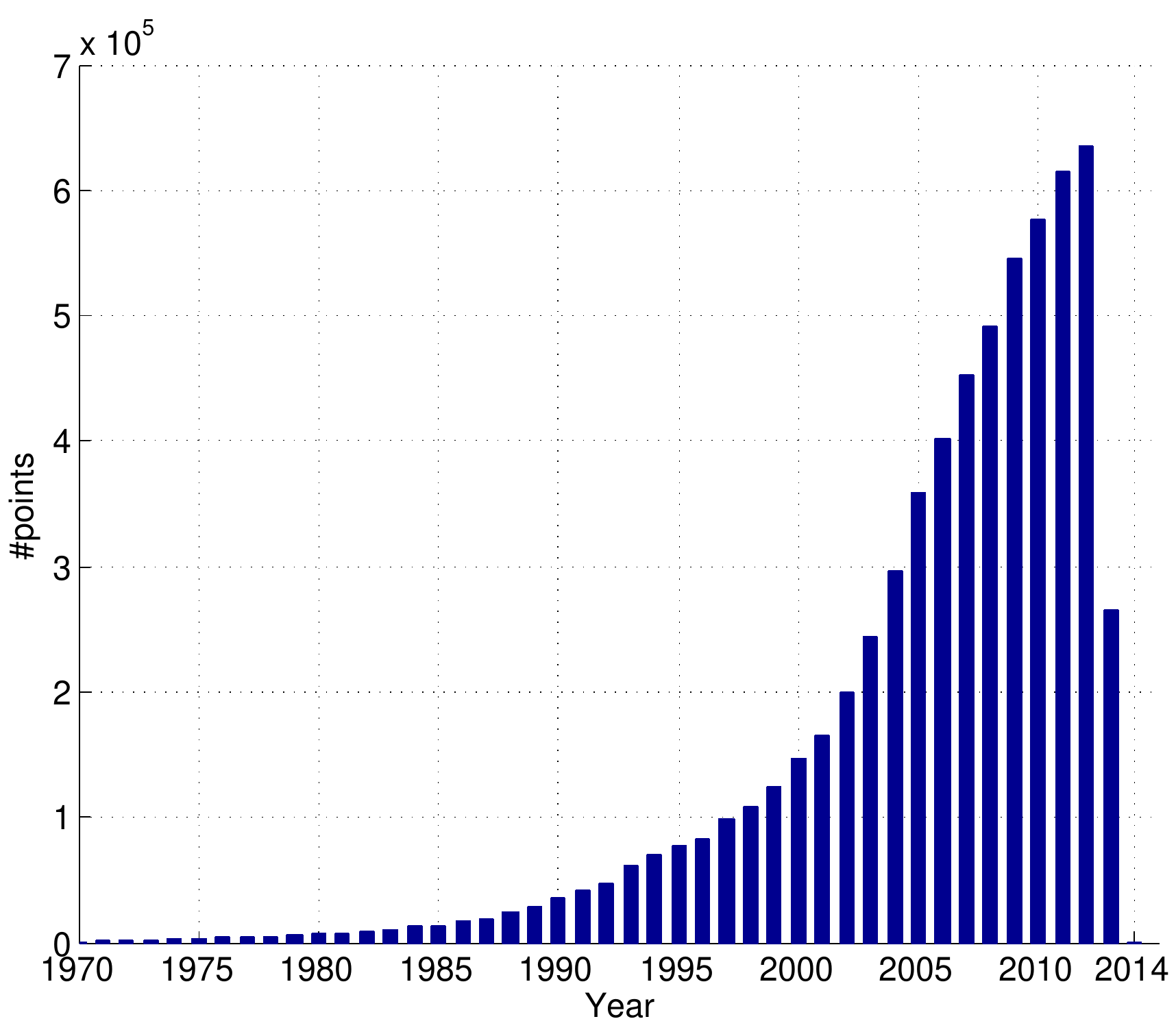}%
&
\includegraphics[width=.29\columnwidth]{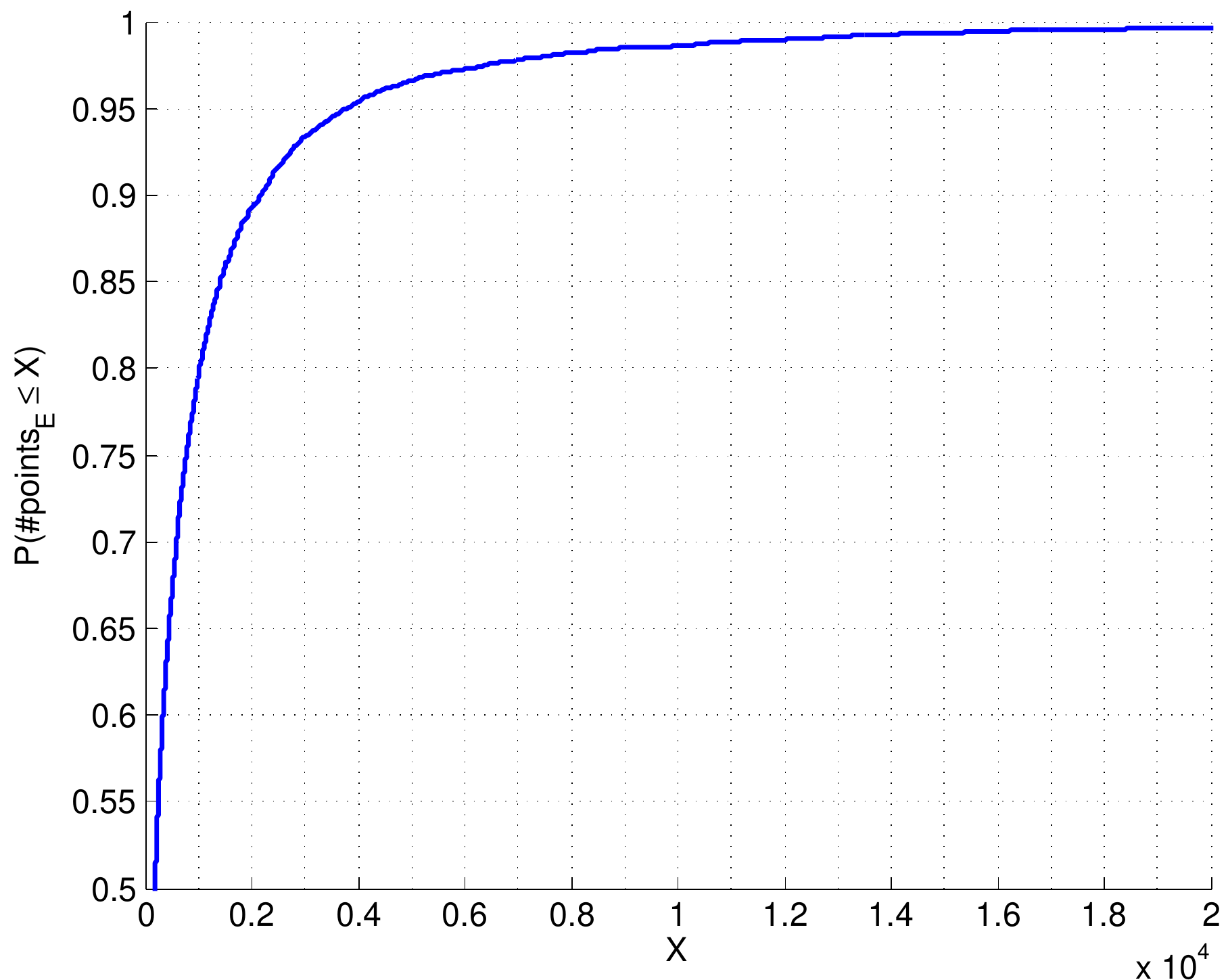}\\
(a) & (b) & (c)\\%
\end{tabular}
\caption{Basics on marginal statistics for DBLP cats data. (a) empirical cumulative distribution of authors that have published $X$ times; (b) distribution of points over the years; (c) empirical cumulative distribution of points with event value $e\in E$ that appears $X$ times.}%
\label{fig:marginal}%
\end{figure}

The DBLP cats data shows skewed marginal distributions (see figure~\ref{fig:marginal}): (a) 80\% of authors have published less than 5 times; (b) Most of the points come from the last 20 years; (c) half the events appear less than 200 times and 80\% less than 1000 times.

\begin{figure}[htbp!]
\centering
\includegraphics[width=\columnwidth]{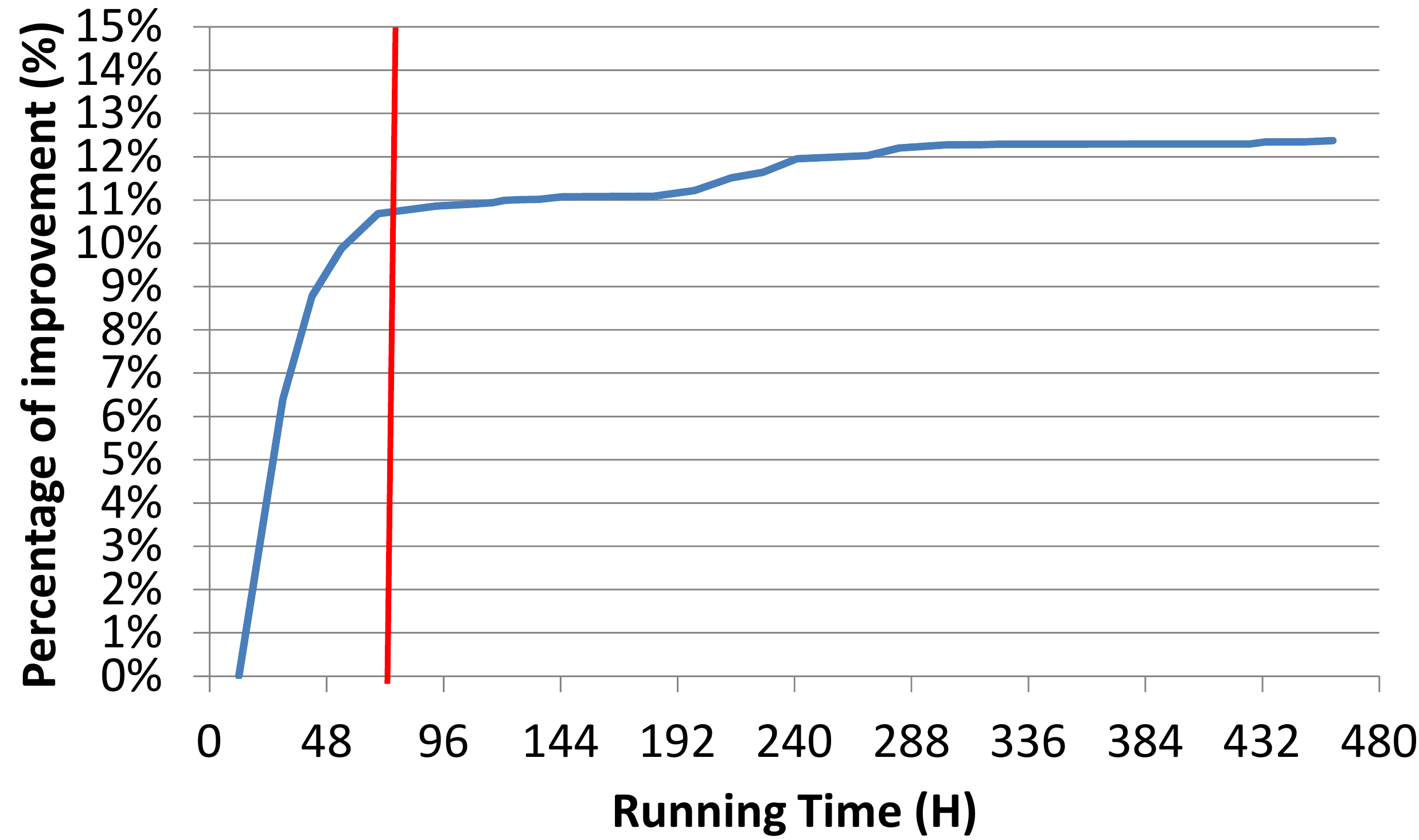}
\caption{Evolution of $cost$ improvement (compared with the first output grid) w.r.t. running time on the whole DBLP cats data set.}%
\label{fig:grid_improvement}%
\end{figure}

\noindent \textbf{Big Picture.} To confirm the scalability and robustness of our approach and to obtain a global picture of DBLP cats data, we apply \textsc{khc} on the whole data -- and to the best of our knowledge, it has never been done. A first grid solution is obtained after 12 hours. Several rounds of optimization allow 12\% improvement of the $cost$ criterion and \textsc{khc} ends after 19 days. Figure~\ref{fig:grid_improvement} relates the evolution of $cost$ improvement (compared with the first output grid solution which is already a ``good'' solution) w.r.t. computational time. We observe that most of the improvement is achieved in the first three days, then the improvement is saturated. The anytime facet of \textsc{khc} allows us to stop before the completion of all rounds of optimization and more generally, it allows the analyst to set the amount of time devoted to the mining phase.

Notice that we have also run \textsc{khc} on smaller versions of DBLP cats data, e.g., with only authors having published 5 times or more and with only events appearing more than 40 times in the data: this version of DBLP is made of 240 thousands of authors and 5K events over the same timeline. Obviously, computational time is smaller (first grid obtained in 7 hours, rounds of optimization offer 8\% of improvement and end in 8 days). Similar high-level results described below can be obtained from both experiments since removed data correspond to the least frequent events and authors (also the least typical) that have the smallest impact on the global data grid structure.\\
We think it is worth waiting several days for computation since DBLP cats data is year-scale, thus potential update of such analysis is needed once a year. Using the whole DBLP cats data set, the final grid $M^{\ast}$ is made of 267 clusters of authors, 4 time intervals and 565 clusters of events (i.e., $6\times 10^5$ cells) whereas the finest grid is made of about $6.8\times 10^{11}$ cells.

\begin{figure*}[htbp!]%
\centering
\includegraphics[width=.6\textwidth]{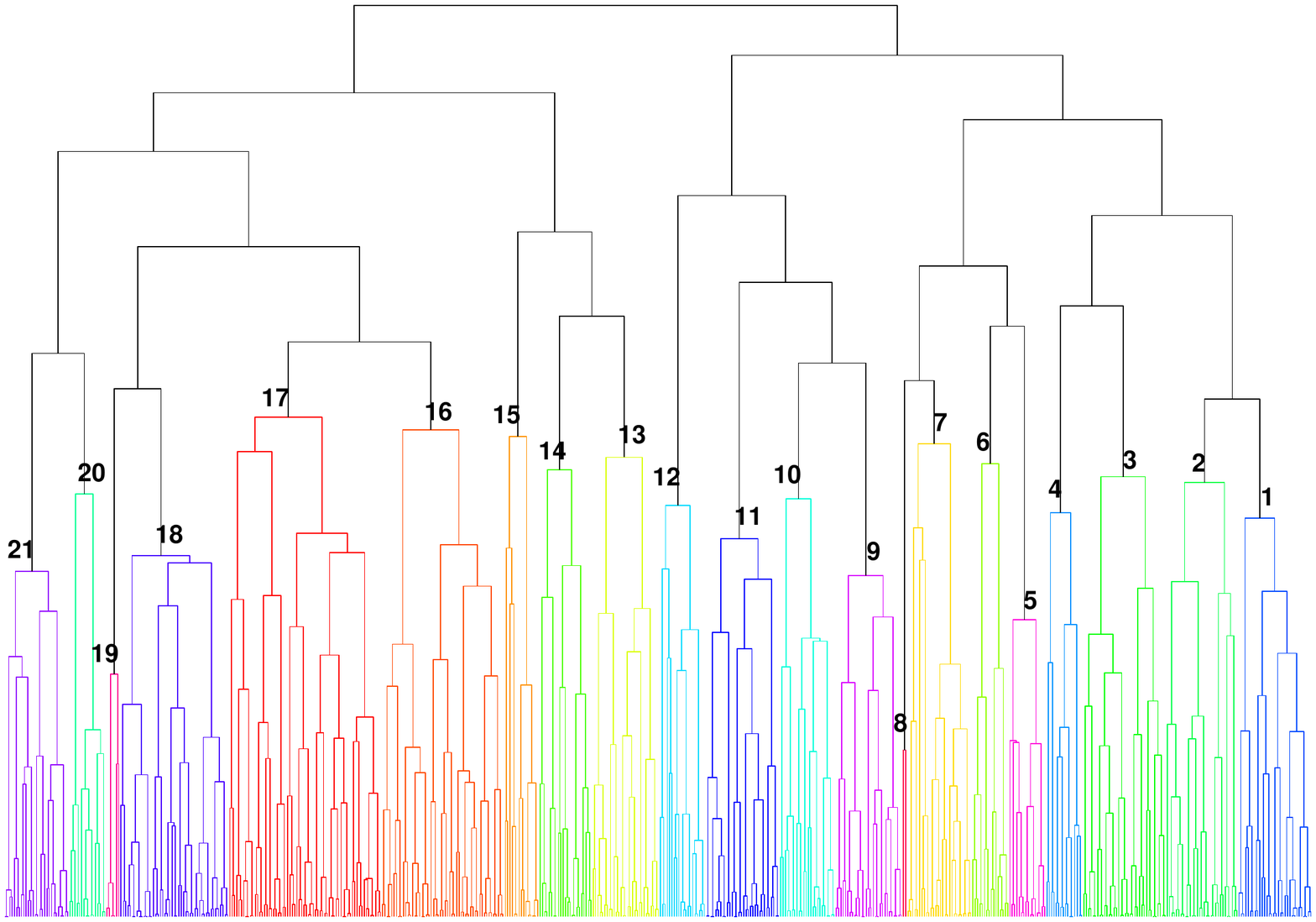}
\caption{Hierarchy of events from DBLP data set. Colored sub-hierarchy at $HierarchicalLevel = 1 - IR(M) = 0.56$, where 21 clusters of events may be labeled easily. 1: Software Engineering; 2: AI, ML, Agents, DB, DM; 3: Educat'lComput, ComputerHumanInteract, VisualInterface; 4: Computat'Linguistics, Info\&TextRetrieval, NLP, Speech-AudioRecog\&Process; 5: FormalMethods, LogicComput\&Prog; 6: Security, Cryptography; 7: DiscreteMaths, Algo, CS\&InformTheory; 8: CoRR\&InformTheory; 9: Real-TimeSystems, Parallel-Distrib-GridComputing; 10: GeneralComput, Systems\&Software; 11: Networks\&Communications; 12: SystIntegration-Config-Design-Architecture; 13: Simulation, AppliedMaths, OperatResearch; 14: MedicalInfo, SignalProcess; 15: ComputBiology, BioInfo; 16: DiverseAppliedCS-1; 17: DiverseAppliedCS-2; 18: AppliedAI\&NeuralNetworks\&FuzzySystems; 19: Robotics; 20: CompGraphics, InfoVisualization; 21: ComputerVision, PatternRec, Multimedia, ImageProcess}%
\label{fig:Alldendrogram}%
\end{figure*}

Using dissimilarity index $\Delta$ and information rate $IR$, from $M^{\ast}$ to $M_{\emptyset}$, we build a hierarchy of clusters or intervals for each dimension. Figure~\ref{fig:Alldendrogram} relates the whole hierarchy for the event dimension and shows how the events (proceedings/journals/others) are organized by topic in DBLP from cats data point of view. Since 565 clusters are hardly interpretable by humans, figure~\ref{fig:Alldendrogram} highlights the hierarchy at $HierarchicalLevel = 1 - IR(M) = 0.56$, i.e., keeping 44\% information from $M^{\ast}$.\\
At this granularity, 21 clusters of events can be interpreted and labeled easily by looking at the most typical event titles of the clusters. Indeed, the found clusters correspond to Computer Science research sub-fields indexed by DBLP. For example, SIGMETRICS, SIGCOMM, INFOCOMM, GLOBECOM, IMC, CoNEXT, IEEE Communications Magazine, \ldots are among the most typical events of terminal clusters under the branch 11 (labeled Networks\&Communications) because recurrent researchers in that field regularly published in these events over the years and not significantly in other fields like e.g., 19: Robotics. Notice also the singularity of cluster 8 which mainly consists of references of Computing Research Repository (CoRR) covering many sub-fields of Computer Science research.

\begin{figure}[htbp!]
\centering
\includegraphics[width=\columnwidth]{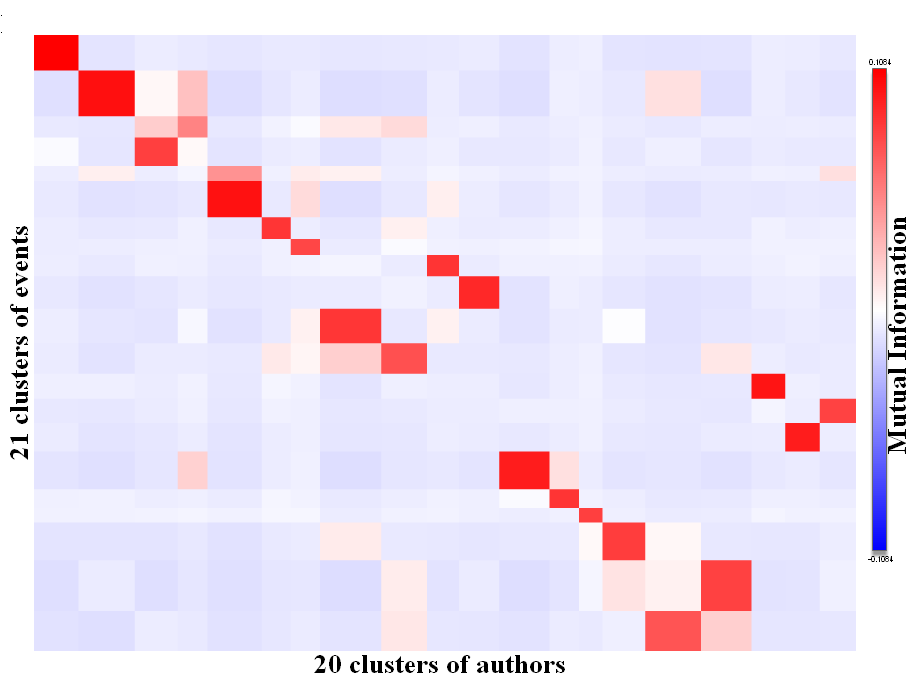}
\caption{CMI visualization for $A\times E$ for the whole time period.}%
\label{fig:AE_matrix}%
\end{figure}

At this scale, the grid is made of 21 clusters of events, 4 time intervals and 20 clusters of authors. Considering the 2D visualization of CMI for Author and Event dimensions (see figure~\ref{fig:AE_matrix}), red cells (i.e., positive CMI relates significant positive interactions) highlight the diagonal 2D-cells of the $A\times E$ matrix with different color intensity depending on the time interval or the whole period considered, while the other light blue cells indicate a deficit of interactions. The diagonal form of interactions between authors and events (actually almost diagonal due to the singular cluster 8 including CoRR) indicates that, at this scale, most of the researchers (grouped in clusters) are active exclusively in a unique sub-field.\\
Another interesting observation may be made when considering the two agglomerate clusters (clusters 1-12 and 13-21 from figure~\ref{fig:Alldendrogram}) at the top the hierarchy: the former mainly relates to fundamental research while the latter is more focused on applied research. 
%
\begin{figure*}[htbp!]
\centering
\includegraphics[width=.6\textwidth]{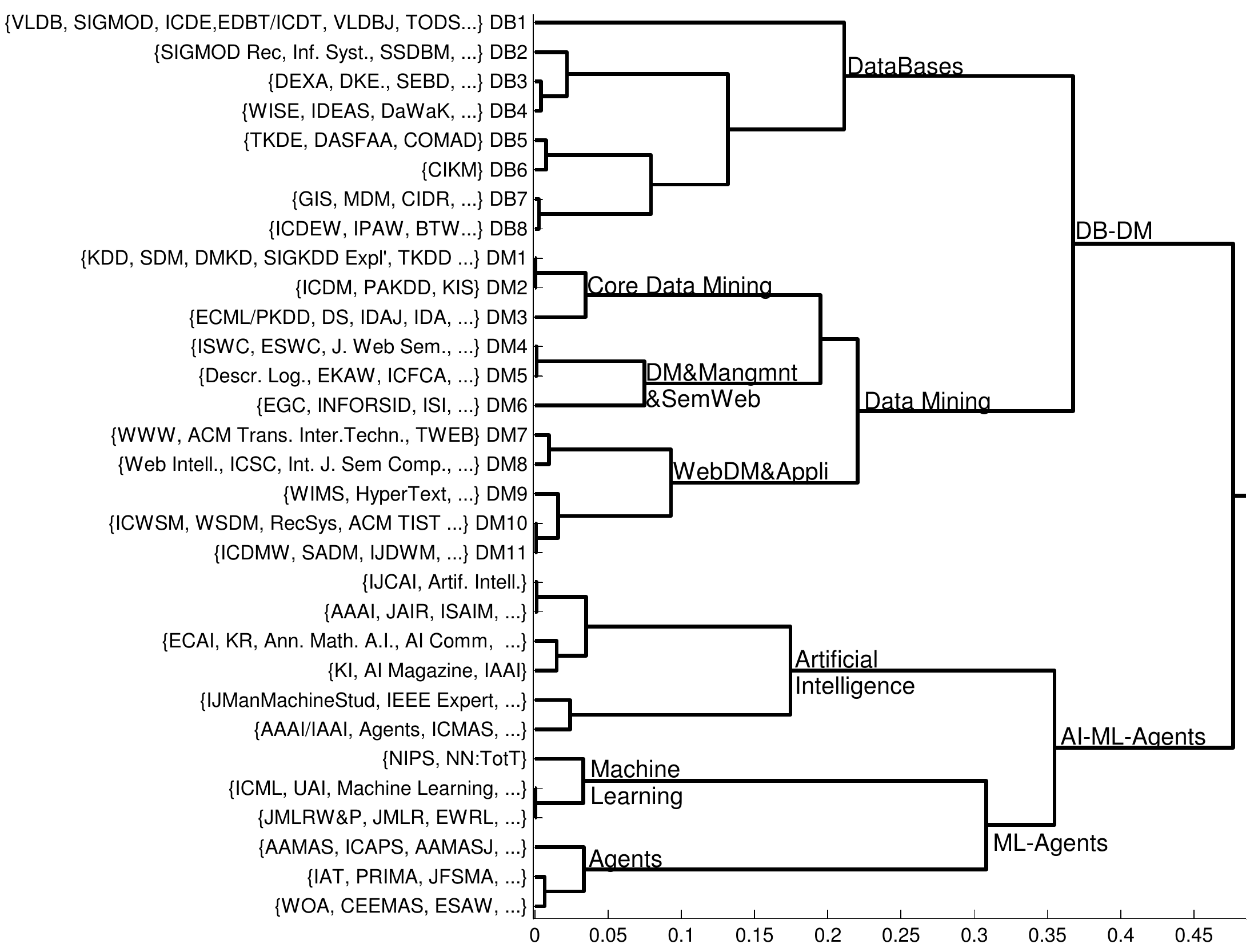}
\caption{Sub-hierarchy corresponding to close but different sub-fields: Machine Learning, Artificial Intelligence, Agents, Data Bases, Data Mining. ($X$-axis: $HierarchicalLevel = 1 - IR(M)$, $Y$-axis: most typical events of terminal clusters).}%
\label{fig:AIdendrogram}%
\end{figure*}

\noindent \textbf{Zoom in the \emph{Data Bases} and \emph{Data Mining} fields.} Figure~\ref{fig:AIdendrogram} details the sub-hierarchy of cluster 2 (AI, ML, Agents, DB, DM) from figure~\ref{fig:Alldendrogram}. For the same reason as above, and even if there exist authors across several sub-fields of cluster 2, Data Bases, Data Mining, Machine Learning, Artificial Intelligence and Agents Systems are recognized as close though different sub-fields -- confirming the intuition; particularly, Data Mining is closer to Data Bases than AI, ML or Agents.

\begin{figure}[htbp!]%
\centering
\begin{tabular}{cc}
Typical Authors & Contrast\\
\includegraphics[width=.33\columnwidth]{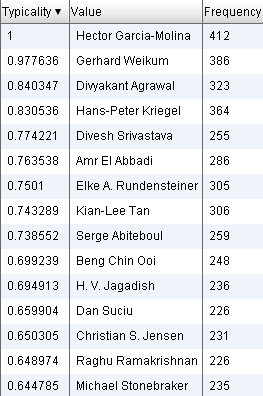}%
&
\includegraphics[width=.6\columnwidth]{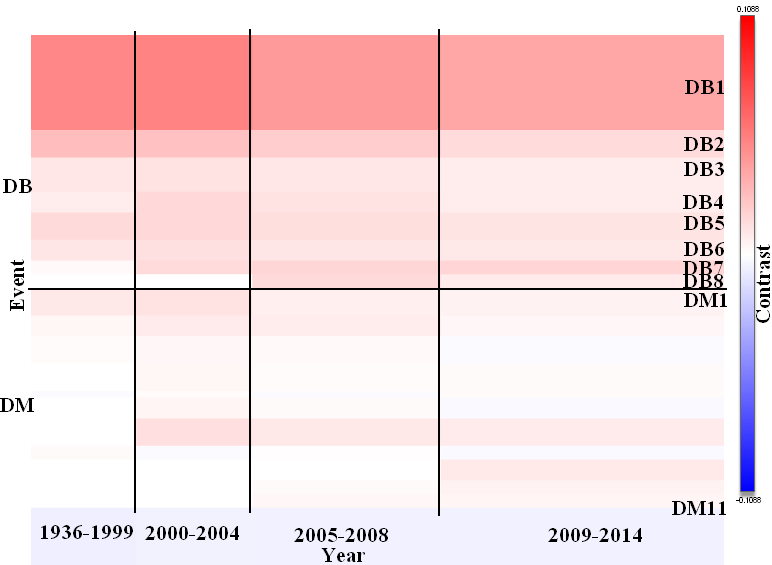}\\
\includegraphics[width=.33\columnwidth]{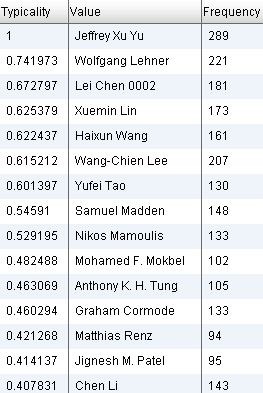}%
&
\includegraphics[width=.6\columnwidth]{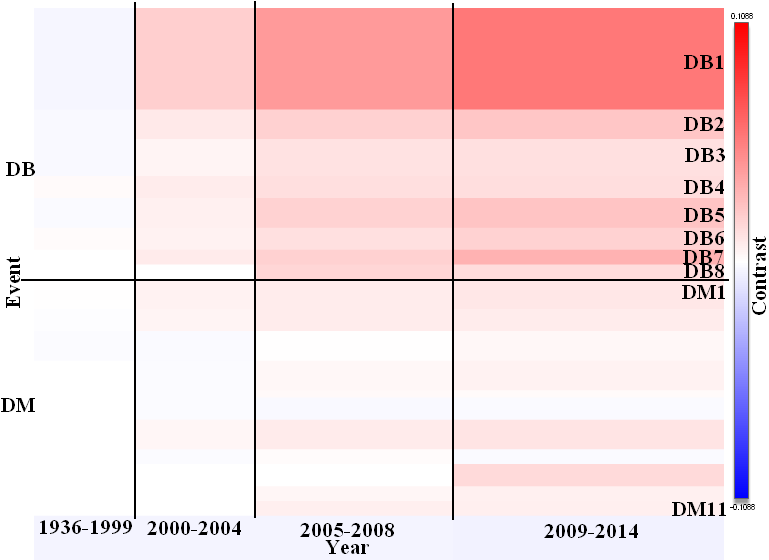}\\
%
%
\end{tabular}
\caption{Visualization of two clusters of authors, their typical authors, and contrast in 2D $T\times E$ view. Positive contrast (red cells) highlights what is characteristic of the cluster. ($DB_i$ and $DM_j$ correspond to terminal clusters highlight in figure~\ref{fig:AIdendrogram}.}%
\label{fig:dblp_clusters}%
\end{figure}

In figure~\ref{fig:dblp_clusters}, we present two terminal clusters of authors who are involved in DB/DM research. Since clusters of authors may contain thousands of authors (as shown by figure~\ref{fig:marginal}, most of them have published less than 5 times) and frequency visualization is not enough to characterize the clusters, we show the 15 most typical authors and what is contrasting in their trajectory of publications over the years w.r.t. the rest of the data.\\
We observe that cluster (a) is made of senior DB researchers (H. Garcia-Molina, D. Weikum, D. Agrawal, \ldots) who are characterized by their activity in DB events in the whole time period with a strong contrast before 2004 in the top DB events. Cluster (b) whose most typical authors are e.g., J. Xu Yu, W. Lehner, \ldots is made of experienced but younger DB researchers: the contrast highlight their activity in DB field in the last ten years.\\
We have also found clusters of authors who are characterized (by contrast) by their activity in (core) Data Mining research (resp. Semantic Web): typical authors of those clusters are well-known experienced researchers like P.S. Yu, J. Han, C. Faloutsos, \ldots (resp. I. Horrocks, S. Staab, W. Neijdl, \ldots) who cover their respective field since its birth. Similar observations can be made for any other sub-fields discovered by the hierarchy of event clusters in figure~\ref{fig:Alldendrogram}.\\
Thus, starting from the big picture provided by the 3D-grid computed by \textsc{khc} on the whole DBLP cats data, we are able to zoom in discovered sub-fields of Computer Science research indexed by DBLP, to obtain the most typical authors of clusters of authors that are involved in the field and to explain the characteristics of their sequence of publications in terms of contrast.

\section{Background and related work}
\label{sec:related}
Since most standard numerical time series clustering algorithms are based on (dis) similarity measures, several (dis) similarity measures have been designed and exploited for categorical time series clustering: e.g., the Levenshtein distance for clustering life courses~\cite{MGR+08}, the discrete Fr\'echet distance for clustering migration data~\cite{PGS+13}, the Compression-based Distance Measure (CDM)~\cite{KLR04}\ldots These measures are adapted to classical clustering and partitioning algorithms; they often require parameter tuning and do not directly offer abilities to interpret and explore the results for time and event variable.

Coclustering methods may be classified into two different branches:
\begin{itemize}
\item coclustering methods for object $\times$ attributes (see pioneering work~\cite{Har72})
\item coclustering methods for two or more attributes like Dhillon et al.~\cite{DMM03}, which is the most related to our work 
\end{itemize}
Dhillon et al.~\cite{DMM03} have proposed an information-theoretic coclustering approach for two discrete random variables: the loss in Mutual Information $MI(X,Y) - MI(X^{\pi_M} , Y^{\pi_M})$ is minimized to obtain a locally-optimal grid with a user-defined mandatory number of clusters for each dimension. 

The Information Bottleneck (IB) method~\cite{TPB99} stems from another information-theoretic paradigm: 
Given the joint probability $P(X,Y)$, IB aims at grouping $X$ into clusters $T$ in order to both compress $X$ and keep as much information as possible about $Y$. IB also minimizes a difference in Mutual Information: $MI(T,X) - \beta MI(T,Y)$, where $\beta$ is a positive Lagrange multiplier. Wang et al.~\cite{WDL10} build upon IB and suggest a coclustering method for two categorical variables.\\
Extending IB for more than two categorical variables, Slonim et al.~\cite{SFT01} have suggested the agglomerative multivariate information bottleneck that allows constructing several interacting systems of clusters simultaneously; the interactions among variables are specified using a Bayesian network structure.

There also exist many research works that suggest solutions for the problem of segmentation of one event sequence: e.g., Kiernan \& Terzi~\cite{KT08,KT09} suggest a parameter-free method for building interpretable summaries (through segmentation) of an event sequence.

Going beyond 2D matrices, recent significant progress has been done in multi-way tensor analysis~\cite{STF06,KS08}. For instance, \cite{MSF+12} suggest a method for mining time-stamped event sequences and effective forecasting of future events, but does not answer to the problem of clustering. Also, \cite{MSF14} explore the problem of mining co-evolving time sequences but is mainly dedicated to numerical sequences. Both recent approaches scale well with the time dimension, but it is not demonstrated that they are effective and efficient on data with many sequences.

As far as we know, there is no method building upon above recent related work and suggesting an effective and efficient solution to large-scale clustering of categorical time series.

%

\section{Conclusion \& discussion}
\label{sec:conclusion}
We have suggested a method for coclustering and exploratory analysis of categorical time series (or temporal event sequences) based on three-dimensional data grid models. The sequence identifiers are grouped into clusters, the time dimension is discretized into intervals and the events are also grouped into clusters -- the whole forming a 3D-grid. The optimal grid (the most probable a posteriori in Bayesian terms) is obtained with an user parameter-free procedure. To exploit the resulting grid, we have suggested \textit{(i)} a dissimilarity index between clusters to select the wanted granularity of the grid while controlling the information loss; \textit{(ii)}, a criterion, namely the typicality, to rank and identify representative values in a cluster; \textit{(iii)} two other criteria stemming from Mutual Information to characterize, interpret and visualize the found clusters. Our insightful findings have been illustrated on both synthetic and real-world data sets.\\
As future work, we plan to extend the approach to supervised cats classification and event forecasting. Looking at another direction, due to the genericity of the approach, we plan to apply \textsc{khc} to other application domains and data sets: actually, we are currently working on the behavior analysis of customer trajectories, where a customer (cats) is defined by his actions on various communication channels (e.g., interactive voice system, after-sales service calls, services/products browsing, storefront shops visits, etc.) over various time periods.\\

%

%
\bibliographystyle{abbrv}
\bibliography{biblio_general}  
\end{document}